
\magnification=\magstep1


\catcode`\@=11


\message{Loading jyTeX fonts...}



\font\vptrm=cmr5
\font\vptmit=cmmi5
\font\vptsy=cmsy5
\font\vptbf=cmbx5

\skewchar\vptmit='177 \skewchar\vptsy='60
\fontdimen16 \vptsy=\the\fontdimen17 \vptsy

\def\vpt{\ifmmode\err@badsizechange\else
     \@mathfontinit
     \textfont0=\vptrm  \scriptfont0=\vptrm  \scriptscriptfont0=\vptrm
     \textfont1=\vptmit \scriptfont1=\vptmit \scriptscriptfont1=\vptmit
     \textfont2=\vptsy  \scriptfont2=\vptsy  \scriptscriptfont2=\vptsy
     \textfont3=\xptex  \scriptfont3=\xptex  \scriptscriptfont3=\xptex
     \textfont\bffam=\vptbf
     \scriptfont\bffam=\vptbf
     \scriptscriptfont\bffam=\vptbf
     \@fontstyleinit
     \def\rm{\vptrm\fam=\z@}%
     \def\bf{\vptbf\fam=\bffam}%
     \def\oldstyle{\vptmit\fam=\@ne}%
     \rm\fi}


\font\viptrm=cmr6
\font\viptmit=cmmi6
\font\viptsy=cmsy6
\font\viptbf=cmbx6

\skewchar\viptmit='177 \skewchar\viptsy='60
\fontdimen16 \viptsy=\the\fontdimen17 \viptsy

\def\vipt{\ifmmode\err@badsizechange\else
     \@mathfontinit
     \textfont0=\viptrm  \scriptfont0=\vptrm  \scriptscriptfont0=\vptrm
     \textfont1=\viptmit \scriptfont1=\vptmit \scriptscriptfont1=\vptmit
     \textfont2=\viptsy  \scriptfont2=\vptsy  \scriptscriptfont2=\vptsy
     \textfont3=\xptex   \scriptfont3=\xptex  \scriptscriptfont3=\xptex
     \textfont\bffam=\viptbf
     \scriptfont\bffam=\vptbf
     \scriptscriptfont\bffam=\vptbf
     \@fontstyleinit
     \def\rm{\viptrm\fam=\z@}%
     \def\bf{\viptbf\fam=\bffam}%
     \def\oldstyle{\viptmit\fam=\@ne}%
     \rm\fi}


\font\viiptrm=cmr7
\font\viiptmit=cmmi7
\font\viiptsy=cmsy7
\font\viiptit=cmti7
\font\viiptbf=cmbx7

\skewchar\viiptmit='177 \skewchar\viiptsy='60
\fontdimen16 \viiptsy=\the\fontdimen17 \viiptsy

\def\viipt{\ifmmode\err@badsizechange\else
     \@mathfontinit
     \textfont0=\viiptrm  \scriptfont0=\vptrm  \scriptscriptfont0=\vptrm
     \textfont1=\viiptmit \scriptfont1=\vptmit \scriptscriptfont1=\vptmit
     \textfont2=\viiptsy  \scriptfont2=\vptsy  \scriptscriptfont2=\vptsy
     \textfont3=\xptex    \scriptfont3=\xptex  \scriptscriptfont3=\xptex
     \textfont\itfam=\viiptit
     \scriptfont\itfam=\viiptit
     \scriptscriptfont\itfam=\viiptit
     \textfont\bffam=\viiptbf
     \scriptfont\bffam=\vptbf
     \scriptscriptfont\bffam=\vptbf
     \@fontstyleinit
     \def\rm{\viiptrm\fam=\z@}%
     \def\it{\viiptit\fam=\itfam}%
     \def\bf{\viiptbf\fam=\bffam}%
     \def\oldstyle{\viiptmit\fam=\@ne}%
     \rm\fi}


\font\viiiptrm=cmr8
\font\viiiptmit=cmmi8
\font\viiiptsy=cmsy8
\font\viiiptit=cmti8
\font\viiiptbf=cmbx8

\skewchar\viiiptmit='177 \skewchar\viiiptsy='60
\fontdimen16 \viiiptsy=\the\fontdimen17 \viiiptsy

\def\viiipt{\ifmmode\err@badsizechange\else
     \@mathfontinit
     \textfont0=\viiiptrm  \scriptfont0=\viptrm  \scriptscriptfont0=\vptrm
     \textfont1=\viiiptmit \scriptfont1=\viptmit \scriptscriptfont1=\vptmit
     \textfont2=\viiiptsy  \scriptfont2=\viptsy  \scriptscriptfont2=\vptsy
     \textfont3=\xptex     \scriptfont3=\xptex   \scriptscriptfont3=\xptex
     \textfont\itfam=\viiiptit
     \scriptfont\itfam=\viiptit
     \scriptscriptfont\itfam=\viiptit
     \textfont\bffam=\viiiptbf
     \scriptfont\bffam=\viptbf
     \scriptscriptfont\bffam=\vptbf
     \@fontstyleinit
     \def\rm{\viiiptrm\fam=\z@}%
     \def\it{\viiiptit\fam=\itfam}%
     \def\bf{\viiiptbf\fam=\bffam}%
     \def\oldstyle{\viiiptmit\fam=\@ne}%
     \rm\fi}


\def\getixpt{%
     \font\ixptrm=cmr9
     \font\ixptmit=cmmi9
     \font\ixptsy=cmsy9
     \font\ixptit=cmti9
     \font\ixptbf=cmbx9
     \skewchar\ixptmit='177 \skewchar\ixptsy='60
     \fontdimen16 \ixptsy=\the\fontdimen17 \ixptsy}

\def\ixpt{\ifmmode\err@badsizechange\else
     \@mathfontinit
     \textfont0=\ixptrm  \scriptfont0=\viiptrm  \scriptscriptfont0=\vptrm
     \textfont1=\ixptmit \scriptfont1=\viiptmit \scriptscriptfont1=\vptmit
     \textfont2=\ixptsy  \scriptfont2=\viiptsy  \scriptscriptfont2=\vptsy
     \textfont3=\xptex   \scriptfont3=\xptex    \scriptscriptfont3=\xptex
     \textfont\itfam=\ixptit
     \scriptfont\itfam=\viiptit
     \scriptscriptfont\itfam=\viiptit
     \textfont\bffam=\ixptbf
     \scriptfont\bffam=\viiptbf
     \scriptscriptfont\bffam=\vptbf
     \@fontstyleinit
     \def\rm{\ixptrm\fam=\z@}%
     \def\it{\ixptit\fam=\itfam}%
     \def\bf{\ixptbf\fam=\bffam}%
     \def\oldstyle{\ixptmit\fam=\@ne}%
     \rm\fi}


\font\xptrm=cmr10
\font\xptmit=cmmi10
\font\xptsy=cmsy10
\font\xptex=cmex10
\font\xptit=cmti10
\font\xptsl=cmsl10
\font\xptbf=cmbx10
\font\xpttt=cmtt10
\font\xptss=cmss10
\font\xptsc=cmcsc10
\font\xptbfs=cmb10
\font\xptbmit=cmmib10

\skewchar\xptmit='177 \skewchar\xptbmit='177 \skewchar\xptsy='60
\fontdimen16 \xptsy=\the\fontdimen17 \xptsy

\def\xpt{\ifmmode\err@badsizechange\else
     \@mathfontinit
     \textfont0=\xptrm  \scriptfont0=\viiptrm  \scriptscriptfont0=\vptrm
     \textfont1=\xptmit \scriptfont1=\viiptmit \scriptscriptfont1=\vptmit
     \textfont2=\xptsy  \scriptfont2=\viiptsy  \scriptscriptfont2=\vptsy
     \textfont3=\xptex  \scriptfont3=\xptex    \scriptscriptfont3=\xptex
     \textfont\itfam=\xptit
     \scriptfont\itfam=\viiptit
     \scriptscriptfont\itfam=\viiptit
     \textfont\bffam=\xptbf
     \scriptfont\bffam=\viiptbf
     \scriptscriptfont\bffam=\vptbf
     \textfont\bfsfam=\xptbfs
     \scriptfont\bfsfam=\viiptbf
     \scriptscriptfont\bfsfam=\vptbf
     \textfont\bmitfam=\xptbmit
     \scriptfont\bmitfam=\viiptmit
     \scriptscriptfont\bmitfam=\vptmit
     \@fontstyleinit
     \def\rm{\xptrm\fam=\z@}%
     \def\it{\xptit\fam=\itfam}%
     \def\sl{\xptsl}%
     \def\bf{\xptbf\fam=\bffam}%
     \def\tt{\xpttt}%
     \def\ss{\xptss}%
     \def\sc{\xptsc}%
     \def\bfs{\xptbfs\fam=\bfsfam}%
     \def\bmit{\fam=\bmitfam}%
     \def\oldstyle{\xptmit\fam=\@ne}%
     \rm\fi}


\def\getxipt{%
     \font\xiptrm=cmr10  scaled\magstephalf
     \font\xiptmit=cmmi10 scaled\magstephalf
     \font\xiptsy=cmsy10 scaled\magstephalf
     \font\xiptex=cmex10 scaled\magstephalf
     \font\xiptit=cmti10 scaled\magstephalf
     \font\xiptsl=cmsl10 scaled\magstephalf
     \font\xiptbf=cmbx10 scaled\magstephalf
     \font\xipttt=cmtt10 scaled\magstephalf
     \font\xiptss=cmss10 scaled\magstephalf
     \skewchar\xiptmit='177 \skewchar\xiptsy='60
     \fontdimen16 \xiptsy=\the\fontdimen17 \xiptsy}

\def\xipt{\ifmmode\err@badsizechange\else
     \@mathfontinit
     \textfont0=\xiptrm  \scriptfont0=\viiiptrm  \scriptscriptfont0=\viptrm
     \textfont1=\xiptmit \scriptfont1=\viiiptmit \scriptscriptfont1=\viptmit
     \textfont2=\xiptsy  \scriptfont2=\viiiptsy  \scriptscriptfont2=\viptsy
     \textfont3=\xiptex  \scriptfont3=\xptex     \scriptscriptfont3=\xptex
     \textfont\itfam=\xiptit
     \scriptfont\itfam=\viiiptit
     \scriptscriptfont\itfam=\viiptit
     \textfont\bffam=\xiptbf
     \scriptfont\bffam=\viiiptbf
     \scriptscriptfont\bffam=\viptbf
     \@fontstyleinit
     \def\rm{\xiptrm\fam=\z@}%
     \def\it{\xiptit\fam=\itfam}%
     \def\sl{\xiptsl}%
     \def\bf{\xiptbf\fam=\bffam}%
     \def\tt{\xipttt}%
     \def\ss{\xiptss}%
     \def\oldstyle{\xiptmit\fam=\@ne}%
     \rm\fi}


\font\xiiptrm=cmr12
\font\xiiptmit=cmmi12
\font\xiiptsy=cmsy10  scaled\magstep1
\font\xiiptex=cmex10  scaled\magstep1
\font\xiiptit=cmti12
\font\xiiptsl=cmsl12
\font\xiiptbf=cmbx12
\font\xiipttt=cmtt12
\font\xiiptss=cmss12
\font\xiiptsc=cmcsc10 scaled\magstep1
\font\xiiptbfs=cmb10  scaled\magstep1
\font\xiiptbmit=cmmib10 scaled\magstep1

\skewchar\xiiptmit='177 \skewchar\xiiptbmit='177 \skewchar\xiiptsy='60
\fontdimen16 \xiiptsy=\the\fontdimen17 \xiiptsy

\def\xiipt{\ifmmode\err@badsizechange\else
     \@mathfontinit
     \textfont0=\xiiptrm  \scriptfont0=\viiiptrm  \scriptscriptfont0=\viptrm
     \textfont1=\xiiptmit \scriptfont1=\viiiptmit \scriptscriptfont1=\viptmit
     \textfont2=\xiiptsy  \scriptfont2=\viiiptsy  \scriptscriptfont2=\viptsy
     \textfont3=\xiiptex  \scriptfont3=\xptex     \scriptscriptfont3=\xptex
     \textfont\itfam=\xiiptit
     \scriptfont\itfam=\viiiptit
     \scriptscriptfont\itfam=\viiptit
     \textfont\bffam=\xiiptbf
     \scriptfont\bffam=\viiiptbf
     \scriptscriptfont\bffam=\viptbf
     \textfont\bfsfam=\xiiptbfs
     \scriptfont\bfsfam=\viiiptbf
     \scriptscriptfont\bfsfam=\viptbf
     \textfont\bmitfam=\xiiptbmit
     \scriptfont\bmitfam=\viiiptmit
     \scriptscriptfont\bmitfam=\viptmit
     \@fontstyleinit
     \def\rm{\xiiptrm\fam=\z@}%
     \def\it{\xiiptit\fam=\itfam}%
     \def\sl{\xiiptsl}%
     \def\bf{\xiiptbf\fam=\bffam}%
     \def\tt{\xiipttt}%
     \def\ss{\xiiptss}%
     \def\sc{\xiiptsc}%
     \def\bfs{\xiiptbfs\fam=\bfsfam}%
     \def\bmit{\fam=\bmitfam}%
     \def\oldstyle{\xiiptmit\fam=\@ne}%
     \rm\fi}


\def\getxiiipt{%
     \font\xiiiptrm=cmr12  scaled\magstephalf
     \font\xiiiptmit=cmmi12 scaled\magstephalf
     \font\xiiiptsy=cmsy9  scaled\magstep2
     \font\xiiiptit=cmti12 scaled\magstephalf
     \font\xiiiptsl=cmsl12 scaled\magstephalf
     \font\xiiiptbf=cmbx12 scaled\magstephalf
     \font\xiiipttt=cmtt12 scaled\magstephalf
     \font\xiiiptss=cmss12 scaled\magstephalf
     \skewchar\xiiiptmit='177 \skewchar\xiiiptsy='60
     \fontdimen16 \xiiiptsy=\the\fontdimen17 \xiiiptsy}

\def\xiiipt{\ifmmode\err@badsizechange\else
     \@mathfontinit
     \textfont0=\xiiiptrm  \scriptfont0=\xptrm  \scriptscriptfont0=\viiptrm
     \textfont1=\xiiiptmit \scriptfont1=\xptmit \scriptscriptfont1=\viiptmit
     \textfont2=\xiiiptsy  \scriptfont2=\xptsy  \scriptscriptfont2=\viiptsy
     \textfont3=\xivptex   \scriptfont3=\xptex  \scriptscriptfont3=\xptex
     \textfont\itfam=\xiiiptit
     \scriptfont\itfam=\xptit
     \scriptscriptfont\itfam=\viiptit
     \textfont\bffam=\xiiiptbf
     \scriptfont\bffam=\xptbf
     \scriptscriptfont\bffam=\viiptbf
     \@fontstyleinit
     \def\rm{\xiiiptrm\fam=\z@}%
     \def\it{\xiiiptit\fam=\itfam}%
     \def\sl{\xiiiptsl}%
     \def\bf{\xiiiptbf\fam=\bffam}%
     \def\tt{\xiiipttt}%
     \def\ss{\xiiiptss}%
     \def\oldstyle{\xiiiptmit\fam=\@ne}%
     \rm\fi}


\font\xivptrm=cmr12   scaled\magstep1
\font\xivptmit=cmmi12  scaled\magstep1
\font\xivptsy=cmsy10  scaled\magstep2
\font\xivptex=cmex10  scaled\magstep2
\font\xivptit=cmti12  scaled\magstep1
\font\xivptsl=cmsl12  scaled\magstep1
\font\xivptbf=cmbx12  scaled\magstep1
\font\xivpttt=cmtt12  scaled\magstep1
\font\xivptss=cmss12  scaled\magstep1
\font\xivptsc=cmcsc10 scaled\magstep2
\font\xivptbfs=cmb10  scaled\magstep2
\font\xivptbmit=cmmib10 scaled\magstep2

\skewchar\xivptmit='177 \skewchar\xivptbmit='177 \skewchar\xivptsy='60
\fontdimen16 \xivptsy=\the\fontdimen17 \xivptsy

\def\xivpt{\ifmmode\err@badsizechange\else
     \@mathfontinit
     \textfont0=\xivptrm  \scriptfont0=\xptrm  \scriptscriptfont0=\viiptrm
     \textfont1=\xivptmit \scriptfont1=\xptmit \scriptscriptfont1=\viiptmit
     \textfont2=\xivptsy  \scriptfont2=\xptsy  \scriptscriptfont2=\viiptsy
     \textfont3=\xivptex  \scriptfont3=\xptex  \scriptscriptfont3=\xptex
     \textfont\itfam=\xivptit
     \scriptfont\itfam=\xptit
     \scriptscriptfont\itfam=\viiptit
     \textfont\bffam=\xivptbf
     \scriptfont\bffam=\xptbf
     \scriptscriptfont\bffam=\viiptbf
     \textfont\bfsfam=\xivptbfs
     \scriptfont\bfsfam=\xptbfs
     \scriptscriptfont\bfsfam=\viiptbf
     \textfont\bmitfam=\xivptbmit
     \scriptfont\bmitfam=\xptbmit
     \scriptscriptfont\bmitfam=\viiptmit
     \@fontstyleinit
     \def\rm{\xivptrm\fam=\z@}%
     \def\it{\xivptit\fam=\itfam}%
     \def\sl{\xivptsl}%
     \def\bf{\xivptbf\fam=\bffam}%
     \def\tt{\xivpttt}%
     \def\ss{\xivptss}%
     \def\sc{\xivptsc}%
     \def\bfs{\xivptbfs\fam=\bfsfam}%
     \def\bmit{\fam=\bmitfam}%
     \def\oldstyle{\xivptmit\fam=\@ne}%
     \rm\fi}


\font\xviiptrm=cmr17
\font\xviiptmit=cmmi12 scaled\magstep2
\font\xviiptsy=cmsy10 scaled\magstep3
\font\xviiptex=cmex10 scaled\magstep3
\font\xviiptit=cmti12 scaled\magstep2
\font\xviiptbf=cmbx12 scaled\magstep2
\font\xviiptbfs=cmb10 scaled\magstep3

\skewchar\xviiptmit='177 \skewchar\xviiptsy='60
\fontdimen16 \xviiptsy=\the\fontdimen17 \xviiptsy

\def\xviipt{\ifmmode\err@badsizechange\else
     \@mathfontinit
     \textfont0=\xviiptrm  \scriptfont0=\xiiptrm  \scriptscriptfont0=\viiiptrm
     \textfont1=\xviiptmit \scriptfont1=\xiiptmit \scriptscriptfont1=\viiiptmit
     \textfont2=\xviiptsy  \scriptfont2=\xiiptsy  \scriptscriptfont2=\viiiptsy
     \textfont3=\xviiptex  \scriptfont3=\xiiptex  \scriptscriptfont3=\xptex
     \textfont\itfam=\xviiptit
     \scriptfont\itfam=\xiiptit
     \scriptscriptfont\itfam=\viiiptit
     \textfont\bffam=\xviiptbf
     \scriptfont\bffam=\xiiptbf
     \scriptscriptfont\bffam=\viiiptbf
     \textfont\bfsfam=\xviiptbfs
     \scriptfont\bfsfam=\xiiptbfs
     \scriptscriptfont\bfsfam=\viiiptbf
     \@fontstyleinit
     \def\rm{\xviiptrm\fam=\z@}%
     \def\it{\xviiptit\fam=\itfam}%
     \def\bf{\xviiptbf\fam=\bffam}%
     \def\bfs{\xviiptbfs\fam=\bfsfam}%
     \def\oldstyle{\xviiptmit\fam=\@ne}%
     \rm\fi}


\font\xxiptrm=cmr17  scaled\magstep1


\def\xxipt{\ifmmode\err@badsizechange\else
     \@mathfontinit
     \@fontstyleinit
     \def\rm{\xxiptrm\fam=\z@}%
     \rm\fi}


\font\xxvptrm=cmr17  scaled\magstep2


\def\xxvpt{\ifmmode\err@badsizechange\else
     \@mathfontinit
     \@fontstyleinit
     \def\rm{\xxvptrm\fam=\z@}%
     \rm\fi}




\message{Loading jyTeX macros...}

\message{modifications to plain.tex,}


\def\newcount{\alloc@0\count\countdef\insc@unt}
\def\newdimen{\alloc@1\dimen\dimendef\insc@unt}
\def\newskip{\alloc@2\skip\skipdef\insc@unt}
\def\newmuskip{\alloc@3\muskip\muskipdef\@cclvi}
\def\newbox{\alloc@4\box\chardef\insc@unt}
\def\newtoks{\alloc@5\toks\toksdef\@cclvi}
\def\newhelp#1#2{\newtoks#1\global#1\expandafter{\csname#2\endcsname}}
\def\newread{\alloc@6\read\chardef\sixt@@n}
\def\newwrite{\alloc@7\write\chardef\sixt@@n}
\def\newfam{\alloc@8\fam\chardef\sixt@@n}
\def\newinsert#1{\global\advance\insc@unt by\m@ne
     \ch@ck0\insc@unt\count
     \ch@ck1\insc@unt\dimen
     \ch@ck2\insc@unt\skip
     \ch@ck4\insc@unt\box
     \allocationnumber=\insc@unt
     \global\chardef#1=\allocationnumber
     \wlog{\string#1=\string\insert\the\allocationnumber}}
\def\newif#1{\count@\escapechar \escapechar\m@ne
     \expandafter\expandafter\expandafter
          \xdef\@if#1{true}{\let\noexpand#1=\noexpand\iftrue}%
     \expandafter\expandafter\expandafter
          \xdef\@if#1{false}{\let\noexpand#1=\noexpand\iffalse}%
     \global\@if#1{false}\escapechar=\count@}


\newlinechar=`\^^J
\overfullrule=0pt




\let\itfam=\undefined

\let\bffam=\undefined

\count18=3


\chardef\sharps="19


\mathchardef\alpha="710B
\mathchardef\beta="710C
\mathchardef\gamma="710D
\mathchardef\delta="710E
\mathchardef\epsilon="710F
\mathchardef\zeta="7110
\mathchardef\eta="7111
\mathchardef\theta="7112
\mathchardef\iota="7113
\mathchardef\kappa="7114
\mathchardef\lambda="7115
\mathchardef\mu="7116
\mathchardef\nu="7117
\mathchardef\xi="7118
\mathchardef\pi="7119
\mathchardef\rho="711A
\mathchardef\sigma="711B
\mathchardef\tau="711C
\mathchardef\upsilon="711D
\mathchardef\phi="711E
\mathchardef\chi="711F
\mathchardef\psi="7120
\mathchardef\omega="7121
\mathchardef\varepsilon="7122
\mathchardef\vartheta="7123
\mathchardef\varpi="7124
\mathchardef\varrho="7125
\mathchardef\varsigma="7126
\mathchardef\varphi="7127
\mathchardef\imath="717B
\mathchardef\jmath="717C
\mathchardef\ell="7160
\mathchardef\wp="717D
\mathchardef\partial="7140
\mathchardef\flat="715B
\mathchardef\natural="715C
\mathchardef\sharp="715D



\def\angle{{\vbox{\ialign{$\m@th\scriptstyle##$\crcr
     \not\mathrel{\mkern14mu}\crcr
     \noalign{\nointerlineskip}
     \mkern2.5mu\leaders\hrule height.34\rp@\hfill\mkern2.5mu\crcr}}}}
\def\vdots{\vbox{\baselineskip4\rp@ \lineskiplimit\z@
     \kern6\rp@\hbox{.}\hbox{.}\hbox{.}}}
\def\ddots{\mathinner{\mkern1mu\raise7\rp@\vbox{\kern7\rp@\hbox{.}}\mkern2mu
     \raise4\rp@\hbox{.}\mkern2mu\raise\rp@\hbox{.}\mkern1mu}}
\def\overrightarrow#1{\vbox{\ialign{##\crcr
     \rightarrowfill\crcr
     \noalign{\kern-\rp@\nointerlineskip}
     $\hfil\displaystyle{#1}\hfil$\crcr}}}
\def\overleftarrow#1{\vbox{\ialign{##\crcr
     \leftarrowfill\crcr
     \noalign{\kern-\rp@\nointerlineskip}
     $\hfil\displaystyle{#1}\hfil$\crcr}}}
\def\overbrace#1{\mathop{\vbox{\ialign{##\crcr
     \noalign{\kern3\rp@}
     \downbracefill\crcr
     \noalign{\kern3\rp@\nointerlineskip}
     $\hfil\displaystyle{#1}\hfil$\crcr}}}\limits}
\def\underbrace#1{\mathop{\vtop{\ialign{##\crcr
     $\hfil\displaystyle{#1}\hfil$\crcr
     \noalign{\kern3\rp@\nointerlineskip}
     \upbracefill\crcr
     \noalign{\kern3\rp@}}}}\limits}
\def\big#1{{\hbox{$\left#1\vbox to8.5\rp@ {}\right.\n@space$}}}
\def\Big#1{{\hbox{$\left#1\vbox to11.5\rp@ {}\right.\n@space$}}}
\def\bigg#1{{\hbox{$\left#1\vbox to14.5\rp@ {}\right.\n@space$}}}
\def\Bigg#1{{\hbox{$\left#1\vbox to17.5\rp@ {}\right.\n@space$}}}
\def\@vereq#1#2{\lower.5\rp@\vbox{\baselineskip\z@skip\lineskip-.5\rp@
     \ialign{$\m@th#1\hfil##\hfil$\crcr#2\crcr=\crcr}}}
\def\rlh@#1{\vcenter{\hbox{\ooalign{\raise2\rp@
     \hbox{$#1\rightharpoonup$}\crcr
     $#1\leftharpoondown$}}}}
\def\bordermatrix#1{\begingroup\m@th
     \setbox\z@\vbox{%
          \def\cr{\crcr\noalign{\kern2\rp@\global\let\cr\endline}}%
          \ialign{$##$\hfil\kern2\rp@\kern\p@renwd
               &\thinspace\hfil$##$\hfil&&\quad\hfil$##$\hfil\crcr
               \omit\strut\hfil\crcr
               \noalign{\kern-\baselineskip}%
               #1\crcr\omit\strut\cr}}%
     \setbox\tw@\vbox{\unvcopy\z@\global\setbox\@ne\lastbox}%
     \setbox\tw@\hbox{\unhbox\@ne\unskip\global\setbox\@ne\lastbox}%
     \setbox\tw@\hbox{$\kern\wd\@ne\kern-\p@renwd\left(\kern-\wd\@ne
          \global\setbox\@ne\vbox{\box\@ne\kern2\rp@}%
          \vcenter{\kern-\ht\@ne\unvbox\z@\kern-\baselineskip}%
          \,\right)$}%
     \null\;\vbox{\kern\ht\@ne\box\tw@}\endgroup}
\def\endinsert{\egroup
     \if@mid\dimen@\ht\z@
          \advance\dimen@\dp\z@
          \advance\dimen@12\rp@
          \advance\dimen@\pagetotal
          \ifdim\dimen@>\pagegoal\@midfalse\p@gefalse\fi
     \fi
     \if@mid\bigskip\box\z@
          \bigbreak
     \else\insert\topins{\penalty100 \splittopskip\z@skip
               \splitmaxdepth\maxdimen\floatingpenalty\z@
               \ifp@ge\dimen@\dp\z@
                    \vbox to\vsize{\unvbox\z@\kern-\dimen@}%
               \else\box\z@\nobreak\bigskip
               \fi}%
     \fi
     \endgroup}


\def\cases#1{\left\{\,\vcenter{\m@th
     \ialign{$##\hfil$&\quad##\hfil\crcr#1\crcr}}\right.}
\def\matrix#1{\null\,\vcenter{\m@th
     \ialign{\hfil$##$\hfil&&\quad\hfil$##$\hfil\crcr
          \mathstrut\crcr
          \noalign{\kern-\baselineskip}
          #1\crcr
          \mathstrut\crcr
          \noalign{\kern-\baselineskip}}}\,}


\newif\ifraggedbottom

\def\raggedbottom{\ifraggedbottom\else
     \advance\topskip by\z@ plus60pt \raggedbottomtrue\fi}%
\def\normalbottom{\ifraggedbottom
     \advance\topskip by\z@ plus-60pt \raggedbottomfalse\fi}

\message{hacks,}


\toksdef\toks@i=1
\toksdef\toks@ii=2


\def\TeX{T\kern-.1667em \lower.5ex \hbox{E}\kern-.125em X\null}
\def\jyTeX{{\leavevmode
     \raise.587ex \hbox{\it\j}\kern-.1em \lower.048ex \hbox{\it y}\kern-.12em
     \TeX}}

\let\then=\iftrue
\def\ifnoarg#1\then{\def\hack@{#1}\ifx\hack@\empty}
\def\ifundefined#1\then{%
     \expandafter\ifx\csname\expandafter\blank\string#1\endcsname\relax}
\def\useif#1\then{\csname#1\endcsname}
\def\usename#1{\csname#1\endcsname}
\def\useafter#1#2{\expandafter#1\csname#2\endcsname}

\long\def\loop#1\repeat{\def\@iterate{#1\expandafter\@iterate\fi}\@iterate
     \let\@iterate=\relax}

\let\TeXend=\end
\def\begin#1{\begingroup\def\@@blockname{#1}\usename{begin#1}}
\def\end#1{\usename{end#1}\def\hack@{#1}%
     \ifx\@@blockname\hack@
          \endgroup
     \else\err@badgroup\hack@\@@blockname
     \fi}
\def\@@blockname{}

\def\defaultoption[#1]#2{%
     \def\hack@{\ifx\hack@ii[\toks@={#2}\else\toks@={#2[#1]}\fi\the\toks@}%
     \futurelet\hack@ii\hack@}

\def\markup#1{\let\@@marksf=\empty
     \ifhmode\edef\@@marksf{\spacefactor=\the\spacefactor\relax}\/\fi
     ${}^{\hbox{\subscriptfonts#1}}$\@@marksf}


\newtoks\shortyear
\newtoks\militaryhour
\newtoks\standardhour
\newtoks\minute
\newtoks\amorpm

\def\settime{\count@=\time\divide\count@ by60
     \militaryhour=\expandafter{\number\count@}%
     {\multiply\count@ by-60 \advance\count@ by\time
          \xdef\hack@{\ifnum\count@<10 0\fi\number\count@}}%
     \minute=\expandafter{\hack@}%
     \ifnum\count@<12
          \amorpm={am}
     \else\amorpm={pm}
          \ifnum\count@>12 \advance\count@ by-12 \fi
     \fi
     \standardhour=\expandafter{\number\count@}%
     \def\hack@19##1##2{\shortyear={##1##2}}%
          \expandafter\hack@\the\year}

\def\monthword#1{%
     \ifcase#1
          $\bullet$\err@badcountervalue{monthword}%
          \or January\or February\or March\or April\or May\or June%
          \or July\or August\or September\or October\or November\or December%
     \else$\bullet$\err@badcountervalue{monthword}%
     \fi}

\def\monthabbr#1{%
     \ifcase#1
          $\bullet$\err@badcountervalue{monthabbr}%
          \or Jan\or Feb\or Mar\or Apr\or May\or Jun%
          \or Jul\or Aug\or Sep\or Oct\or Nov\or Dec%
     \else$\bullet$\err@badcountervalue{monthabbr}%
     \fi}

\def\militarytime{\the\militaryhour:\the\minute}
\def\standardtime{\the\standardhour:\the\minute}


\def\@setnumstyle#1#2{\expandafter\global\expandafter\expandafter
     \expandafter\let\expandafter\expandafter
     \csname @\expandafter\blank\string#1style\endcsname
     \csname#2\endcsname}
\def\numstyle#1{\usename{@\expandafter\blank\string#1style}#1}
\def\ifblank#1\then{\useafter\ifx{@\expandafter\blank\string#1}\blank}

\def\blank#1{}

\def\Roman#1{\expandafter\uppercase\expandafter{\romannumeral#1}}
\def\alphabetic#1{%
     \ifcase#1
          $\bullet$\err@badcountervalue{alphabetic}%
          \or a\or b\or c\or d\or e\or f\or g\or h\or i\or j\or k\or l\or m%
          \or n\or o\or p\or q\or r\or s\or t\or u\or v\or w\or x\or y\or z%
     \else$\bullet$\err@badcountervalue{alphabetic}%
     \fi}
\def\Alphabetic#1{\expandafter\uppercase\expandafter{\alphabetic{#1}}}
\def\symbols#1{%
     \ifcase#1
          $\bullet$\err@badcountervalue{symbols}%
          \or*\or\dag\or\ddag\or\S\or$\|$%
          \or**\or\dag\dag\or\ddag\ddag\or\S\S\or$\|\|$%
     \else$\bullet$\err@badcountervalue{symbols}%
     \fi}


\catcode`\^^?=13 \def^^?{\relax}

\def\trimleading#1\to#2{\edef#2{#1}%
     \expandafter\@trimleading\expandafter#2#2^^?^^?}
\def\@trimleading#1#2#3^^?{\ifx#2^^?\def#1{}\else\def#1{#2#3}\fi}

\def\trimtrailing#1\to#2{\edef#2{#1}%
     \expandafter\@trimtrailing\expandafter#2#2^^? ^^?\relax}
\def\@trimtrailing#1#2 ^^?#3{\ifx#3\relax\toks@={}%
     \else\def#1{#2}\toks@={\trimtrailing#1\to#1}\fi
     \the\toks@}

\def\trim#1\to#2{\trimleading#1\to#2\trimtrailing#2\to#2}

\catcode`\^^?=15


\long\def\additemL#1\to#2{\toks@={\^^\{#1}}\toks@ii=\expandafter{#2}%
     \xdef#2{\the\toks@\the\toks@ii}}

\long\def\additemR#1\to#2{\toks@={\^^\{#1}}\toks@ii=\expandafter{#2}%
     \xdef#2{\the\toks@ii\the\toks@}}

\def\getitemL#1\to#2{\expandafter\@getitemL#1\hack@#1#2}
\def\@getitemL\^^\#1#2\hack@#3#4{\def#4{#1}\def#3{#2}}

\message{font macros,}


\newdimen\rp@
\newcount\@@sizeindex \@@sizeindex=0
\newcount\@@factori
\newcount\@@factorii
\newcount\@@factoriii
\newcount\@@factoriv

\countdef\maxfam=18
\newfam\itfam
\newfam\bffam
\newfam\bfsfam
\newfam\bmitfam

\def\@mathfontinit{\count@=4
     \loop\textfont\count@=\nullfont
          \scriptfont\count@=\nullfont
          \scriptscriptfont\count@=\nullfont
          \ifnum\count@<\maxfam\advance\count@ by\@ne
     \repeat}

\def\@fontstyleinit{%
     \def\it{\err@fontnotavailable\it}%
     \def\bf{\err@fontnotavailable\bf}%
     \def\bfs{\err@bfstobf}%
     \def\bmit{\err@fontnotavailable\bmit}%
     \def\sc{\err@fontnotavailable\sc}%
     \def\sl{\err@sltoit}%
     \def\ss{\err@fontnotavailable\ss}%
     \def\tt{\err@fontnotavailable\tt}}

\def\@parameterinit#1{\rm\rp@=.1em \@getscaling{#1}%
     \let\^^\=\@doscaling\scalingskipslist
     \setbox\strutbox=\hbox{\vrule
          height.708\baselineskip depth.292\baselineskip width\z@}}

\def\@getfactor#1#2#3#4{\@@factori=#1 \@@factorii=#2
     \@@factoriii=#3 \@@factoriv=#4}

\def\@getscaling#1{\count@=#1 \advance\count@ by-\@@sizeindex\@@sizeindex=#1
     \ifnum\count@<0
          \let\@mulordiv=\divide
          \let\@divormul=\multiply
          \multiply\count@ by\m@ne
     \else\let\@mulordiv=\multiply
          \let\@divormul=\divide
     \fi
     \edef\@@scratcha{\ifcase\count@                {1}{1}{1}{1}\or
          {1}{7}{23}{3}\or     {2}{5}{3}{1}\or      {9}{89}{13}{1}\or
          {6}{25}{6}{1}\or     {8}{71}{14}{1}\or    {6}{25}{36}{5}\or
          {1}{7}{53}{4}\or     {12}{125}{108}{5}\or {3}{14}{53}{5}\or
          {6}{41}{17}{1}\or    {13}{31}{13}{2}\or   {9}{107}{71}{2}\or
          {11}{139}{124}{3}\or {1}{6}{43}{2}\or     {10}{107}{42}{1}\or
          {1}{5}{43}{2}\or     {5}{69}{65}{1}\or    {11}{97}{91}{2}\fi}%
     \expandafter\@getfactor\@@scratcha}

\def\@doscaling#1{\@mulordiv#1by\@@factori\@divormul#1by\@@factorii
     \@mulordiv#1by\@@factoriii\@divormul#1by\@@factoriv}


\newskip\headskip
\newskip\footskip

\def\typesize=#1pt{\count@=#1 \advance\count@ by-10
     \ifcase\count@
          \@setsizex\or\err@badtypesize\or
          \@setsizexii\or\err@badtypesize\or
          \@setsizexiv
     \else\err@badtypesize
     \fi}

\def\@setsizex{\getixpt
     \def\subsubscriptfonts{\vpt}%
          \def\subsubscriptsize{\vpt\@parameterinit{-8}}%
     \def\subscriptfonts{\viipt}\def\subscriptsize{\viipt\@parameterinit{-4}}%
     \def\footnotefonts{\viiipt}\def\footnotesize{\viiipt\@parameterinit{-2}}%
     \def\smallfonts{\ixpt}\def\smallsize{\ixpt\@parameterinit{-1}}%
     \def\normalfonts{\xpt}\def\normalsize{\xpt\@parameterinit{0}}%
     \def\bigfonts{\xiipt}\def\bigsize{\xiipt\@parameterinit{2}}%
     \def\Bigfonts{\xivpt}\def\Bigsize{\xivpt\@parameterinit{4}}%
     \def\biggfonts{\xviipt}\def\biggsize{\xviipt\@parameterinit{6}}%
     \def\Biggfonts{\xxipt}\def\Biggsize{\xxipt\@parameterinit{8}}%
     \def\tinyfonts{\vpt}\def\tinysize{\vpt\@parameterinit{-8}}%
     \def\HUGEFONTS{\xxvpt}\def\HUGESIZE{\xxvpt\@parameterinit{10}}%
     \normalsize\fixedskipslist}

\def\@setsizexii{\getxipt
     \def\subsubscriptfonts{\vipt}%
          \def\subsubscriptsize{\vipt\@parameterinit{-6}}%
     \def\subscriptfonts{\viiipt}%
          \def\subscriptsize{\viiipt\@parameterinit{-2}}%
     \def\footnotefonts{\xpt}\def\footnotesize{\xpt\@parameterinit{0}}%
     \def\smallfonts{\xipt}\def\smallsize{\xipt\@parameterinit{1}}%
     \def\normalfonts{\xiipt}\def\normalsize{\xiipt\@parameterinit{2}}%
     \def\bigfonts{\xivpt}\def\bigsize{\xivpt\@parameterinit{4}}%
     \def\Bigfonts{\xviipt}\def\Bigsize{\xviipt\@parameterinit{6}}%
     \def\biggfonts{\xxipt}\def\biggsize{\xxipt\@parameterinit{8}}%
     \def\Biggfonts{\xxvpt}\def\Biggsize{\xxvpt\@parameterinit{10}}%
     \def\tinyfonts{\vpt}\def\tinysize{\vpt\@parameterinit{-8}}%
     \def\HUGEFONTS{\xxvpt}\def\HUGESIZE{\xxvpt\@parameterinit{10}}%
     \normalsize\fixedskipslist}

\def\@setsizexiv{\getxiiipt
     \def\subsubscriptfonts{\viipt}%
          \def\subsubscriptsize{\viipt\@parameterinit{-4}}%
     \def\subscriptfonts{\xpt}\def\subscriptsize{\xpt\@parameterinit{0}}%
     \def\footnotefonts{\xiipt}\def\footnotesize{\xiipt\@parameterinit{2}}%
     \def\smallfonts{\xiiipt}\def\smallsize{\xiiipt\@parameterinit{3}}%
     \def\normalfonts{\xivpt}\def\normalsize{\xivpt\@parameterinit{4}}%
     \def\bigfonts{\xviipt}\def\bigsize{\xviipt\@parameterinit{6}}%
     \def\Bigfonts{\xxipt}\def\Bigsize{\xxipt\@parameterinit{8}}%
     \def\biggfonts{\xxvpt}\def\biggsize{\xxvpt\@parameterinit{10}}%
     \def\Biggfonts{\err@sizetoolarge\Biggfonts\HUGEFONTS}%
          \def\Biggsize{\err@sizetoolarge\Biggsize\HUGESIZE}%
     \def\tinyfonts{\vpt}\def\tinysize{\vpt\@parameterinit{-8}}%
     \def\HUGEFONTS{\xxvpt}\def\HUGESIZE{\xxvpt\@parameterinit{10}}%
     \normalsize\fixedskipslist}

\def\subsubscriptfonts{\vpt} \def\subsubscriptsize{\vpt\@parameterinit{-8}}
\def\subscriptfonts{\viipt}  \def\subscriptsize{\viipt\@parameterinit{-4}}
\def\footnotefonts{\viiipt}  \def\footnotesize{\viiipt\@parameterinit{-2}}
\def\smallfonts{\err@sizenotavailable\smallfonts}
                             \def\smallsize{\ixpt\@parameterinit{-1}}
\def\normalfonts{\xpt}       \def\normalsize{\xpt\@parameterinit{0}}
\def\bigfonts{\xiipt}        \def\bigsize{\xiipt\@parameterinit{2}}
\def\Bigfonts{\xivpt}        \def\Bigsize{\xivpt\@parameterinit{4}}
\def\biggfonts{\xviipt}      \def\biggsize{\xviipt\@parameterinit{6}}
\def\Biggfonts{\xxipt}       \def\Biggsize{\xxipt\@parameterinit{8}}
\def\tinyfonts{\vpt}         \def\tinysize{\vpt\@parameterinit{-8}}
\def\HUGEFONTS{\xxvpt}       \def\HUGESIZE{\xxvpt\@parameterinit{10}}

\message{document layout,}


\newtoks\everyoutput \everyoutput={}
\newdimen\depthofpage
\newcount\pagenum \pagenum=0

\newdimen\oddtopmargin  \newdimen\eventopmargin
\newdimen\oddleftmargin \newdimen\evenleftmargin
\newtoks\oddhead        \newtoks\evenhead
\newtoks\oddfoot        \newtoks\evenfoot

\def\topmargin{\afterassignment\@seteventop\oddtopmargin}
\def\leftmargin{\afterassignment\@setevenleft\oddleftmargin}
\def\head{\afterassignment\@setevenhead\oddhead}
\def\foot{\afterassignment\@setevenfoot\oddfoot}

\def\@seteventop{\eventopmargin=\oddtopmargin}
\def\@setevenleft{\evenleftmargin=\oddleftmargin}
\def\@setevenhead{\evenhead=\oddhead}
\def\@setevenfoot{\evenfoot=\oddfoot}

\def\pagenumstyle#1{\@setnumstyle\pagenum{#1}}

\newif\ifdraft
\def\draft{\drafttrue\leftmargin=.5in \overfullrule=5pt }

\def\outputstyle#1{\global\expandafter\let\expandafter
          \@outputstyle\csname#1output\endcsname
     \usename{#1setup}}

\output={\@outputstyle}

\def\normaloutput{\the\everyoutput
     \global\advance\pagenum by\@ne
     \ifodd\pagenum
          \voffset=\oddtopmargin \hoffset=\oddleftmargin
     \else\voffset=\eventopmargin \hoffset=\evenleftmargin
     \fi
     \advance\voffset by-1in  \advance\hoffset by-1in
     \count0=\pagenum
     \expandafter\shipout\pagebox
     \ifnum\outputpenalty>-\@MM\else\dosupereject\fi}

\newdimen\fullhsize
\newbox\leftpage
\newcount\leftpagenum
\newcount\outputpagenum \outputpagenum=0
\let\leftorright=L

\def\twoupoutput{\the\everyoutput
     \global\advance\pagenum by\@ne
     \if L\leftorright
          \global\setbox\leftpage=\leftline{\pagebox}%
          \global\leftpagenum=\pagenum
          \global\let\leftorright=R%
     \else\global\advance\outputpagenum by\@ne
          \ifodd\outputpagenum
               \voffset=\oddtopmargin \hoffset=\oddleftmargin
          \else\voffset=\eventopmargin \hoffset=\evenleftmargin
          \fi
          \advance\voffset by-1in  \advance\hoffset by-1in
          \count0=\leftpagenum \count1=\pagenum
          \shipout\vbox{\hbox to\fullhsize
               {\box\leftpage\hfil\leftline{\pagebox}}}%
          \global\let\leftorright=L%
     \fi
     \ifnum\outputpenalty>-\@MM
     \else\dosupereject
          \if R\leftorright
               \globaldefs=\@ne\head={\hfil}\foot={\hfil}\globaldefs=\z@
               \null\newpage
          \fi
     \fi}

\def\pagebox{\vbox{\makeheadline\pagebody\makefootline}}

\def\makeheadline{%
     \vbox to\z@{\baselinestretch=\@m
          \vskip\topskip\vskip-.708\baselineskip\vskip-\headskip
          \line{\vbox to\ht\strutbox{}%
               \ifodd\pagenum\the\oddhead\else\the\evenhead\fi}%
          \vss}%
     \nointerlineskip}

\def\pagebody{\vbox to\vsize{%
     \boxmaxdepth\maxdepth
     \ifvoid\topins\else\unvbox\topins\fi
     \depthofpage=\dp255
     \unvbox255
     \ifraggedbottom\kern-\depthofpage\vfil\fi
     \ifvoid\footins
     \else\vskip\skip\footins
          \footnoterule
          \unvbox\footins
          \vskip-\footnoteskip
     \fi}}

\def\makefootline{\baselineskip=\footskip
     \line{\ifodd\pagenum\the\oddfoot\else\the\evenfoot\fi}}


\newskip\abovechapterskip
\newskip\belowchapterskip
\newskip\abovesectionskip
\newskip\belowsectionskip
\newskip\abovesubsectionskip
\newskip\belowsubsectionskip

\def\chapterstyle#1{\global\expandafter\let\expandafter\@chapterstyle
     \csname#1text\endcsname}
\def\sectionstyle#1{\global\expandafter\let\expandafter\@sectionstyle
     \csname#1text\endcsname}
\def\subsectionstyle#1{\global\expandafter\let\expandafter\@subsectionstyle
     \csname#1text\endcsname}

\def\chapter#1{%
     \ifdim\lastskip=17sp \else\chapterbreak\vskip\abovechapterskip\fi
     \@chapterstyle{\ifblank\chapternumstyle\then
          \else\newchapternum=\next\chapternumformat\ \fi#1}%
     \nobreak\vskip\belowchapterskip\vskip17sp }

\def\section#1{%
     \ifdim\lastskip=17sp \else\sectionbreak\vskip\abovesectionskip\fi
     \@sectionstyle{\ifblank\sectionnumstyle\then
          \else\newsectionnum=\next\sectionnumformat\ \fi#1}%
     \nobreak\vskip\belowsectionskip\vskip17sp }

\def\subsection#1{%
     \ifdim\lastskip=17sp \else\subsectionbreak\vskip\abovesubsectionskip\fi
     \@subsectionstyle{\ifblank\subsectionnumstyle\then
          \else\newsubsectionnum=\next\subsectionnumformat\ \fi#1}%
     \nobreak\vskip\belowsubsectionskip\vskip17sp }


\let\TeXunderline=\underline
\let\TeXoverline=\overline
\def\underline#1{\relax\ifmmode\TeXunderline{#1}\else
     $\TeXunderline{\hbox{#1}}$\fi}
\def\overline#1{\relax\ifmmode\TeXoverline{#1}\else
     $\TeXoverline{\hbox{#1}}$\fi}

\def\baselinestretch{\afterassignment\@baselinestretch\count@}
\def\@baselinestretch{\baselineskip=\normalbaselineskip
     \divide\baselineskip by\@m\baselineskip=\count@\baselineskip
     \setbox\strutbox=\hbox{\vrule
          height.708\baselineskip depth.292\baselineskip width\z@}%
     \bigskipamount=\the\baselineskip
          plus.25\baselineskip minus.25\baselineskip
     \medskipamount=.5\baselineskip
          plus.125\baselineskip minus.125\baselineskip
     \smallskipamount=.25\baselineskip
          plus.0625\baselineskip minus.0625\baselineskip}

\def\\{\ifhmode\ifnum\lastpenalty=-\@M\else\hfil\penalty-\@M\fi\fi
     \ignorespaces}
\def\newpage{\vfil\break}

\def\lefttext#1{\par{\@text\leftskip=\z@\rightskip=\centering
     \noindent#1\par}}
\def\righttext#1{\par{\@text\leftskip=\centering\rightskip=\z@
     \noindent#1\par}}
\def\centertext#1{\par{\@text\leftskip=\centering\rightskip=\centering
     \noindent#1\par}}
\def\@text{\parindent=\z@ \parfillskip=\z@ \everypar={}%
     \spaceskip=.3333em \xspaceskip=.5em
     \def\\{\ifhmode\ifnum\lastpenalty=-\@M\else\penalty-\@M\fi\fi
          \ignorespaces}}

\def\beginleft{\par\@text\leftskip=\z@ \rightskip=\centering}
     
\def\beginright{\par\@text\leftskip=\centering\rightskip=\z@ }
     
\def\begincenter{\par\@text\leftskip=\centering\rightskip=\centering}

\def\beginnarrow{\defaultoption[\parindent]\@beginnarrow}
\def\@beginnarrow[#1]{\par\advance\leftskip by#1\advance\rightskip by#1}

\begingroup
\catcode`\[=1 \catcode`\{=11
\gdef\beginignore[\endgroup\bgroup
     \catcode`\e=0 \catcode`\\=12 \catcode`\{=11 \catcode`\f=12 \let\or=\relax
     \let\nd{ignor=\fi \let\}=\egroup
     \iffalse}
\endgroup

\long\def\marginnote#1{\leavevmode
     \edef\@marginsf{\spacefactor=\the\spacefactor\relax}%
     \ifdraft\strut\vadjust{%
          \hbox to\z@{\hskip\hsize\hskip.1in
               \vbox to\z@{\vskip-\dp\strutbox
                    \marginnoteformat
                    \vskip-\ht\strutbox
                    \noindent\strut#1\par
                    \vss}%
               \hss}}%
     \fi
     \@marginsf}


\newtoks\everybye \everybye={\par\vfil}
\outer\def\bye{\the\everybye
     \footnotecheck
     \prelabelcheck
     \streamcheck
     \supereject
     \TeXend}

\message{footnotes,}

\newcount\footnotenum \footnotenum=0
\newskip\footnoteskip
\let\@footnotelist=\empty

\def\footnotenumstyle#1{\@setnumstyle\footnotenum{#1}%
     \useafter\ifx{@footnotenumstyle}\symbols
          \global\let\@footup=\empty
     \else\global\let\@footup=\markup
     \fi}

\def\footnote{\footnotecheck\defaultoption[]\@footnote}
\def\@footnote[#1]{\@footnotemark[#1]\@footnotetext}

\def\footnotemark{\defaultoption[]\@footnotemark}
\def\@footnotemark[#1]{\let\@footsf=\empty
     \ifhmode\edef\@footsf{\spacefactor=\the\spacefactor\relax}\/\fi
     \ifnoarg#1\then
          \global\advance\footnotenum by\@ne
          \@footup{\footnotenumformat}%
          \edef\@@foota{\footnotenum=\the\footnotenum\relax}%
          \expandafter\additemR\expandafter\@footup\expandafter
               {\@@foota\footnotenumformat}\to\@footnotelist
          \global\let\@footnotelist=\@footnotelist
     \else\markup{#1}%
          \additemR\markup{#1}\to\@footnotelist
          \global\let\@footnotelist=\@footnotelist
     \fi
     \@footsf}

\def\footnotetext{%
     \ifx\@footnotelist\empty\err@extrafootnotetext\else\@footnotetext\fi}
\def\@footnotetext{%
     \getitemL\@footnotelist\to\@@foota
     \global\let\@footnotelist=\@footnotelist
     \insert\footins\bgroup
     \footnoteformat
     \splittopskip=\ht\strutbox\splitmaxdepth=\dp\strutbox
     \interlinepenalty=\interfootnotelinepenalty\floatingpenalty=\@MM
     \noindent\llap{\@@foota}\strut
     \bgroup\aftergroup\@footnoteend
     \let\@@scratcha=}
\def\@footnoteend{\strut\par\vskip\footnoteskip\egroup}

\def\footnoterule{\normalfonts
     \kern-.3em \hrule width2in height.04em \kern .26em }

\def\footnotecheck{%
     \ifx\@footnotelist\empty
     \else\err@extrafootnotemark
          \global\let\@footnotelist=\empty
     \fi}

\message{labels,}

\let\@@labeldef=\xdef
\newif\if@labelfile
\newwrite\@labelfile
\let\@prelabellist=\empty

\def\label#1#2{\trim#1\to\@@labarg\edef\@@labtext{#2}%
     \edef\@@labname{lab@\@@labarg}%
     \useafter\ifundefined\@@labname\then\else\@yeslab\fi
     \useafter\@@labeldef\@@labname{#2}%
     \ifstreaming
          \expandafter\toks@\expandafter\expandafter\expandafter
               {\csname\@@labname\endcsname}%
          \immediate\write\streamout{\noexpand\label{\@@labarg}{\the\toks@}}%
     \fi}
\def\@yeslab{%
     \useafter\ifundefined{if\@@labname}\then
          \err@labelredef\@@labarg
     \else\useif{if\@@labname}\then
               \err@labelredef\@@labarg
          \else\global\usename{\@@labname true}%
               \useafter\ifundefined{pre\@@labname}\then
               \else\useafter\ifx{pre\@@labname}\@@labtext
                    \else\err@badlabelmatch\@@labarg
                    \fi
               \fi
               \if@labelfile
               \else\global\@labelfiletrue
                    \immediate\write\sixt@@n{--> Creating file \jobname.lab}%
                    \immediate\openout\@labelfile=\jobname.lab
               \fi
               \immediate\write\@labelfile
                    {\noexpand\prelabel{\@@labarg}{\@@labtext}}%
          \fi
     \fi}

\def\putlab#1{\trim#1\to\@@labarg\edef\@@labname{lab@\@@labarg}%
     \useafter\ifundefined\@@labname\then\@nolab\else\usename\@@labname\fi}
\def\@nolab{%
     \useafter\ifundefined{pre\@@labname}\then
          \undefinedlabelformat
          \err@needlabel\@@labarg
          \useafter\xdef\@@labname{\undefinedlabelformat}%
     \else\usename{pre\@@labname}%
          \useafter\xdef\@@labname{\usename{pre\@@labname}}%
     \fi
     \useafter\newif{if\@@labname}%
     \expandafter\additemR\@@labarg\to\@prelabellist}

\def\prelabel#1{\useafter\gdef{prelab@#1}}

\def\ifundefinedlabel#1\then{%
     \expandafter\ifx\csname lab@#1\endcsname\relax}
\def\useiflab#1\then{\csname iflab@#1\endcsname}

\def\prelabelcheck{{%
     \def\^^\##1{\useiflab{##1}\then\else\err@undefinedlabel{##1}\fi}%
     \@prelabellist}}

\message{equation numbering,}

\newcount\chapternum
\newcount\sectionnum
\newcount\subsectionnum
\newcount\equationnum
\newcount\subequationnum
\newcount\figurenum
\newcount\subfigurenum
\newcount\tablenum
\newcount\subtablenum

\newif\if@subeqncount
\newif\if@subfigcount
\newif\if@subtblcount

\def\newchapternum{\newsectionnum=\z@\@resetnum\chapternum}
\def\newsectionnum{\newsubsectionnum=\z@\@resetnum\sectionnum}
\def\newsubsectionnum{\newequationnum=\z@\newfigurenum=\z@\newtablenum=\z@
     \@resetnum\subsectionnum}
\def\newequationnum{\newsubequationnum=\z@\@resetnum\equationnum}
\def\newsubequationnum{\@resetnum\subequationnum}
\def\newfigurenum{\newsubfigurenum=\z@\@resetnum\figurenum}
\def\newsubfigurenum{\@resetnum\subfigurenum}
\def\newtablenum{\newsubtablenum=\z@\@resetnum\tablenum}
\def\newsubtablenum{\@resetnum\subtablenum}

\def\@resetnum#1{\global\advance#1by1 \edef\next{\the#1\relax}\global#1}

\newchapternum=0

\def\chapternumstyle#1{\@setnumstyle\chapternum{#1}}
\def\sectionnumstyle#1{\@setnumstyle\sectionnum{#1}}
\def\subsectionnumstyle#1{\@setnumstyle\subsectionnum{#1}}
\def\equationnumstyle#1{\@setnumstyle\equationnum{#1}}
\def\subequationnumstyle#1{\@setnumstyle\subequationnum{#1}%
     \ifblank\subequationnumstyle\then\global\@subeqncountfalse\fi
     \ignorespaces}
\def\figurenumstyle#1{\@setnumstyle\figurenum{#1}}
\def\subfigurenumstyle#1{\@setnumstyle\subfigurenum{#1}%
     \ifblank\subfigurenumstyle\then\global\@subfigcountfalse\fi
     \ignorespaces}
\def\tablenumstyle#1{\@setnumstyle\tablenum{#1}}
\def\subtablenumstyle#1{\@setnumstyle\subtablenum{#1}%
     \ifblank\subtablenumstyle\then\global\@subtblcountfalse\fi
     \ignorespaces}

\def\eqnlabel#1{%
     \if@subeqncount
          \newsubequationnum=\next
     \else\newequationnum=\next
          \ifblank\subequationnumstyle\then
          \else\global\@subeqncounttrue
               \newsubequationnum=\@ne
          \fi
     \fi
     \label{#1}{\puteqnformat}(\puteqn{#1})%
     \ifdraft\rlap{\hskip.1in{\tt#1}}\fi}

\let\puteqn=\putlab

\def\equation#1#2{\useafter\gdef{eqn@#1}{#2\eqno\eqnlabel{#1}}}
\def\Equation#1{\useafter\gdef{eqn@#1}}

\def\putequation#1{\useafter\ifundefined{eqn@#1}\then
     \err@undefinedeqn{#1}\else\usename{eqn@#1}\fi}

\def\eqnseriesstyle#1{\gdef\@eqnseriesstyle{#1}}
\def\begineqnseries{\subequationnumstyle{\@eqnseriesstyle}%
     \defaultoption[]\@begineqnseries}
\def\@begineqnseries[#1]{\edef\@@eqnname{#1}}
\def\endeqnseries{\subequationnumstyle{blank}%
     \expandafter\ifnoarg\@@eqnname\then
     \else\label\@@eqnname{\puteqnformat}%
     \fi
     \aftergroup\ignorespaces}

\def\figlabel#1{%
     \if@subfigcount
          \newsubfigurenum=\next
     \else\newfigurenum=\next
          \ifblank\subfigurenumstyle\then
          \else\global\@subfigcounttrue
               \newsubfigurenum=\@ne
          \fi
     \fi
     \label{#1}{\putfigformat}\putfig{#1}%
     {\def\marginnoteformat{\tt}\marginnote{#1}}}

\let\putfig=\putlab

\def\figseriesstyle#1{\gdef\@figseriesstyle{#1}}
\def\beginfigseries{\subfigurenumstyle{\@figseriesstyle}%
     \defaultoption[]\@beginfigseries}
\def\@beginfigseries[#1]{\edef\@@figname{#1}}
\def\endfigseries{\subfigurenumstyle{blank}%
     \expandafter\ifnoarg\@@figname\then
     \else\label\@@figname{\putfigformat}%
     \fi
     \aftergroup\ignorespaces}

\def\tbllabel#1{%
     \if@subtblcount
          \newsubtablenum=\next
     \else\newtablenum=\next
          \ifblank\subtablenumstyle\then
          \else\global\@subtblcounttrue
               \newsubtablenum=\@ne
          \fi
     \fi
     \label{#1}{\puttblformat}\puttbl{#1}%
     {\def\marginnoteformat{\tt}\marginnote{#1}}}

\let\puttbl=\putlab

\def\tblseriesstyle#1{\gdef\@tblseriesstyle{#1}}
\def\begintblseries{\subtablenumstyle{\@tblseriesstyle}%
     \defaultoption[]\@begintblseries}
\def\@begintblseries[#1]{\edef\@@tblname{#1}}
\def\endtblseries{\subtablenumstyle{blank}%
     \expandafter\ifnoarg\@@tblname\then
     \else\label\@@tblname{\puttblformat}%
     \fi
     \aftergroup\ignorespaces}

\message{reference numbering,}

\newcount\referencenum \referencenum=0
\newcount\@@prerefcount \@@prerefcount=0
\newcount\@@thisref
\newcount\@@lastref
\newcount\@@loopref
\newcount\@@refseq
\newdimen\refnumindent
\let\@undefreflist=\empty

\def\referencenumstyle#1{\@setnumstyle\referencenum{#1}}

\def\referencestyle#1{\usename{@ref#1}}

\def\@refsequential{%
     \gdef\@refpredef##1{\global\advance\referencenum by\@ne
          \let\^^\=0\label{##1}{\^^\{\the\referencenum}}%
          \useafter\gdef{ref@\the\referencenum}{{##1}{\undefinedlabelformat}}}%
     \gdef\@reference##1##2{%
          \ifundefinedlabel##1\then
          \else\def\^^\####1{\global\@@thisref=####1\relax}\putlab{##1}%
               \useafter\gdef{ref@\the\@@thisref}{{##1}{##2}}%
          \fi}%
     \gdef\endputreferences{%
          \loop\ifnum\@@loopref<\referencenum
                    \advance\@@loopref by\@ne
                    \expandafter\expandafter\expandafter\@printreference
                         \csname ref@\the\@@loopref\endcsname
          \repeat
          \par}}

\def\@refpreordered{%
     \gdef\@refpredef##1{\global\advance\referencenum by\@ne
          \additemR##1\to\@undefreflist}%
     \gdef\@reference##1##2{%
          \ifundefinedlabel##1\then
          \else\global\advance\@@loopref by\@ne
               {\let\^^\=0\label{##1}{\^^\{\the\@@loopref}}}%
               \@printreference{##1}{##2}%
          \fi}
     \gdef\endputreferences{%
          \def\^^\####1{\useiflab{####1}\then
               \else\reference{####1}{\undefinedlabelformat}\fi}%
          \@undefreflist
          \par}}

\def\beginprereferences{\par
     \def\reference##1##2{\global\advance\referencenum by1\@ne
          \let\^^\=0\label{##1}{\^^\{\the\referencenum}}%
          \useafter\gdef{ref@\the\referencenum}{{##1}{##2}}}}
\def\endprereferences{\global\@@prerefcount=\the\referencenum\par}

\def\beginputreferences{\par
     \refnumindent=\z@\@@loopref=\z@
     \loop\ifnum\@@loopref<\referencenum
               \advance\@@loopref by\@ne
               \setbox\z@=\hbox{\referencenum=\@@loopref
                    \referencenumformat\enskip}%
               \ifdim\wd\z@>\refnumindent\refnumindent=\wd\z@\fi
     \repeat
     \putreferenceformat
     \@@loopref=\z@
     \loop\ifnum\@@loopref<\@@prerefcount
               \advance\@@loopref by\@ne
               \expandafter\expandafter\expandafter\@printreference
                    \csname ref@\the\@@loopref\endcsname
     \repeat
     \let\reference=\@reference}

\def\@printreference#1#2{\ifx#2\undefinedlabelformat\err@undefinedref{#1}\fi
     \noindent\ifdraft\rlap{\hskip\hsize\hskip.1in \tt#1}\fi
     \llap{\referencenum=\@@loopref\referencenumformat\enskip}#2\par}

\def\reference#1#2{{\par\refnumindent=\z@\putreferenceformat\noindent#2\par}}

\def\putref#1{\trim#1\to\@@refarg
     \expandafter\ifnoarg\@@refarg\then
          \toks@={\relax}%
     \else\@@lastref=-\@m\def\@@refsep{}\def\@more{\@nextref}%
          \toks@={\@nextref#1,,}%
     \fi\the\toks@}
\def\@nextref#1,{\trim#1\to\@@refarg
     \expandafter\ifnoarg\@@refarg\then
          \let\@more=\relax
     \else\ifundefinedlabel\@@refarg\then
               \expandafter\@refpredef\expandafter{\@@refarg}%
          \fi
          \def\^^\##1{\global\@@thisref=##1\relax}%
          \global\@@thisref=\m@ne
          \setbox\z@=\hbox{\putlab\@@refarg}%
     \fi
     \advance\@@lastref by\@ne
     \ifnum\@@lastref=\@@thisref\advance\@@refseq by\@ne\else\@@refseq=\@ne\fi
     \ifnum\@@lastref<\z@
     \else\ifnum\@@refseq<\thr@@
               \@@refsep\def\@@refsep{,}%
               \ifnum\@@lastref>\z@
                    \advance\@@lastref by\m@ne
                    {\referencenum=\@@lastref\putrefformat}%
               \else\undefinedlabelformat
               \fi
          \else\def\@@refsep{--}%
          \fi
     \fi
     \@@lastref=\@@thisref
     \@more}

\message{streaming,}

\newif\ifstreaming

\def\streamto{\defaultoption[\jobname]\@streamto}
\def\@streamto[#1]{\global\streamingtrue
     \immediate\write\sixt@@n{--> Streaming to #1.str}%
     \newwrite\streamout\immediate\openout\streamout=#1.str }

\def\streamfrom{\defaultoption[\jobname]\@streamfrom}
\def\@streamfrom[#1]{\newread\streamin\openin\streamin=#1.str
     \ifeof\streamin
          \expandafter\err@nostream\expandafter{#1.str}%
     \else\immediate\write\sixt@@n{--> Streaming from #1.str}%
          \let\@@labeldef=\gdef
          \ifstreaming
               \edef\@elc{\endlinechar=\the\endlinechar}%
               \endlinechar=\m@ne
               \loop\read\streamin to\@@scratcha
                    \ifeof\streamin
                         \streamingfalse
                    \else\toks@=\expandafter{\@@scratcha}%
                         \immediate\write\streamout{\the\toks@}%
                    \fi
                    \ifstreaming
               \repeat
               \@elc
               \input #1.str
               \streamingtrue
          \else\input #1.str
          \fi
          \let\@@labeldef=\xdef
     \fi}

\def\streamcheck{\ifstreaming
     \immediate\write\streamout{\pagenum=\the\pagenum}%
     \immediate\write\streamout{\footnotenum=\the\footnotenum}%
     \immediate\write\streamout{\referencenum=\the\referencenum}%
     \immediate\write\streamout{\chapternum=\the\chapternum}%
     \immediate\write\streamout{\sectionnum=\the\sectionnum}%
     \immediate\write\streamout{\subsectionnum=\the\subsectionnum}%
     \immediate\write\streamout{\equationnum=\the\equationnum}%
     \immediate\write\streamout{\subequationnum=\the\subequationnum}%
     \immediate\write\streamout{\figurenum=\the\figurenum}%
     \immediate\write\streamout{\subfigurenum=\the\subfigurenum}%
     \immediate\write\streamout{\tablenum=\the\tablenum}%
     \immediate\write\streamout{\subtablenum=\the\subtablenum}%
     \immediate\closeout\streamout
     \fi}


\def\err@badtypesize{%
     \errhelp={The limited availability of certain fonts requires^^J%
          that the base type size be 10pt, 12pt, or 14pt.^^J}%
     \errmessage{--> Illegal base type size}}

\def\err@badsizechange{\immediate\write\sixt@@n
     {--> Size change not allowed in math mode, ignored}}

\def\err@sizetoolarge#1{\immediate\write\sixt@@n
     {--> \noexpand#1 too big, substituting HUGE}}

\def\err@sizenotavailable#1{\immediate\write\sixt@@n
     {--> Size not available, \noexpand#1 ignored}}

\def\err@fontnotavailable#1{\immediate\write\sixt@@n
     {--> Font not available, \noexpand#1 ignored}}

\def\err@sltoit{\immediate\write\sixt@@n
     {--> Style \noexpand\sl not available, substituting \noexpand\it}%
     \it}

\def\err@bfstobf{\immediate\write\sixt@@n
     {--> Style \noexpand\bfs not available, substituting \noexpand\bf}%
     \bf}

\def\err@badgroup#1#2{%
     \errhelp={The block you have just tried to close was not the one^^J%
          most recently opened.^^J}%
     \errmessage{--> \noexpand\end{#1} doesn't match \noexpand\begin{#2}}}

\def\err@badcountervalue#1{\immediate\write\sixt@@n
     {--> Counter (#1) out of bounds}}

\def\err@extrafootnotemark{\immediate\write\sixt@@n
     {--> \noexpand\footnotemark command
          has no corresponding \noexpand\footnotetext}}

\def\err@extrafootnotetext{%
     \errhelp{You have given a \noexpand\footnotetext command without first
          specifying^^Ja \noexpand\footnotemark.^^J}%
     \errmessage{--> \noexpand\footnotetext command has no corresponding
          \noexpand\footnotemark}}

\def\err@labelredef#1{\immediate\write\sixt@@n
     {--> Label "#1" redefined}}

\def\err@badlabelmatch#1{\immediate\write\sixt@@n
     {--> Definition of label "#1" doesn't match value in \jobname.lab}}

\def\err@needlabel#1{\immediate\write\sixt@@n
     {--> Label "#1" cited before its definition}}

\def\err@undefinedlabel#1{\immediate\write\sixt@@n
     {--> Label "#1" cited but never defined}}

\def\err@undefinedeqn#1{\immediate\write\sixt@@n
     {--> Equation "#1" not defined}}

\def\err@undefinedref#1{\immediate\write\sixt@@n
     {--> Reference "#1" not defined}}

\def\err@nostream#1{%
     \errhelp={You have tried to input a stream file that doesn't exist.^^J}%
     \errmessage{--> Stream file #1 not found}}

\message{jyTeX initialization}

\everyjob{\immediate\write16{--> jyTeX version \fmtversion}%
     \edef\@@jobname{\jobname}%
     \edef\jobname{\@@jobname}%
     \settime
     \openin0=\jobname.lab
     \ifeof0
     \else\closein0
          \immediate\write16{--> Getting labels from file \jobname.lab}%
          \input\jobname.lab
     \fi}


\def\fixedskipslist{%
     \^^\{\topskip}%
     \^^\{\splittopskip}%
     \^^\{\maxdepth}%
     \^^\{\skip\topins}%
     \^^\{\skip\footins}%
     \^^\{\headskip}%
     \^^\{\footskip}}

\def\scalingskipslist{%
     \^^\{\p@renwd}%
     \^^\{\delimitershortfall}%
     \^^\{\nulldelimiterspace}%
     \^^\{\scriptspace}%
     \^^\{\jot}%
     \^^\{\normalbaselineskip}%
     \^^\{\normallineskip}%
     \^^\{\normallineskiplimit}%
     \^^\{\baselineskip}%
     \^^\{\lineskip}%
     \^^\{\lineskiplimit}%
     \^^\{\bigskipamount}%
     \^^\{\medskipamount}%
     \^^\{\smallskipamount}%
     \^^\{\parskip}%
     \^^\{\parindent}%
     \^^\{\abovedisplayskip}%
     \^^\{\belowdisplayskip}%
     \^^\{\abovedisplayshortskip}%
     \^^\{\belowdisplayshortskip}%
     \^^\{\abovechapterskip}%
     \^^\{\belowchapterskip}%
     \^^\{\abovesectionskip}%
     \^^\{\belowsectionskip}%
     \^^\{\abovesubsectionskip}%
     \^^\{\belowsubsectionskip}}


\def\twoupsetup{
     \topmargin=.75in
     \leftmargin=.5in
     \vsize=6.9in
     \hsize=4.75in
     \fullhsize=10in
     \let\draft=\relax}

\outputstyle{normal}                             

\def\marginnoteformat{\subscriptsize             
     \hsize=1in \baselinestretch=1000 \everypar={}%
     \tolerance=5000 \hbadness=5000 \parskip=0pt \parindent=0pt
     \leftskip=0pt \rightskip=0pt \raggedright}

\head={\ifdraft\normalfonts\it\hfil DRAFT\hfil   
     \llap{\number\day\ \monthword\month\ \militarytime}\else\hfil\fi}
\foot={\hfil\normalfonts\numstyle\pagenum\hfil}  

\normalbaselineskip=12pt                         
\normallineskip=0pt                              
\normallineskiplimit=0pt                         
\normalbaselines                                 

\topskip=.85\baselineskip
\splittopskip=\topskip
\headskip=2\baselineskip
\footskip=\headskip

\pagenumstyle{arabic}                            

\parskip=0pt                                     
\parindent=20pt                                  

\baselinestretch=1000                            


\chapterstyle{left}                              
\chapternumstyle{blank}                          
\def\chapterbreak{\newpage}                      
\abovechapterskip=0pt                            
\belowchapterskip=1.5\baselineskip               
     plus.38\baselineskip minus.38\baselineskip
\def\chapternumformat{\numstyle\chapternum.}     

\sectionstyle{left}                              
\sectionnumstyle{blank}                          
\def\sectionbreak{\vskip0pt plus4\baselineskip\penalty-100
     \vskip0pt plus-4\baselineskip}              
\abovesectionskip=1.5\baselineskip               
     plus.38\baselineskip minus.38\baselineskip
\belowsectionskip=\the\baselineskip              
     plus.25\baselineskip minus.25\baselineskip
\def\sectionnumformat{
     \ifblank\chapternumstyle\then\else\numstyle\chapternum.\fi
     \numstyle\sectionnum.}

\subsectionstyle{left}                           
\subsectionnumstyle{blank}                       
\def\subsectionbreak{\vskip0pt plus4\baselineskip\penalty-100
     \vskip0pt plus-4\baselineskip}              
\abovesubsectionskip=\the\baselineskip           
     plus.25\baselineskip minus.25\baselineskip
\belowsubsectionskip=.75\baselineskip            
     plus.19\baselineskip minus.19\baselineskip
\def\subsectionnumformat{
     \ifblank\chapternumstyle\then\else\numstyle\chapternum.\fi
     \ifblank\sectionnumstyle\then\else\numstyle\sectionnum.\fi
     \numstyle\subsectionnum.}


\footnotenumstyle{symbols}                       
\footnoteskip=0pt                                
\def\footnotenumformat{\numstyle\footnotenum}    
\def\footnoteformat{\footnotesize                
     \everypar={}\parskip=0pt \parfillskip=0pt plus1fil
     \leftskip=1em \rightskip=0pt
     \spaceskip=0pt \xspaceskip=0pt
     \def\\{\ifhmode\ifnum\lastpenalty=-10000
          \else\hfil\penalty-10000 \fi\fi\ignorespaces}}


\def\undefinedlabelformat{$\bullet$}             


\equationnumstyle{arabic}                        
\subequationnumstyle{blank}                      
\figurenumstyle{arabic}                          
\subfigurenumstyle{blank}                        
\tablenumstyle{arabic}                           
\subtablenumstyle{blank}                         

\eqnseriesstyle{alphabetic}                      
\figseriesstyle{alphabetic}                      
\tblseriesstyle{alphabetic}                      

\def\puteqnformat{\hbox{
     \ifblank\chapternumstyle\then\else\numstyle\chapternum.\fi
     \ifblank\sectionnumstyle\then\else\numstyle\sectionnum.\fi
     \ifblank\subsectionnumstyle\then\else\numstyle\subsectionnum.\fi
     \numstyle\equationnum
     \numstyle\subequationnum}}
\def\putfigformat{\hbox{
     \ifblank\chapternumstyle\then\else\numstyle\chapternum.\fi
     \ifblank\sectionnumstyle\then\else\numstyle\sectionnum.\fi
     \ifblank\subsectionnumstyle\then\else\numstyle\subsectionnum.\fi
     \numstyle\figurenum
     \numstyle\subfigurenum}}
\def\puttblformat{\hbox{
     \ifblank\chapternumstyle\then\else\numstyle\chapternum.\fi
     \ifblank\sectionnumstyle\then\else\numstyle\sectionnum.\fi
     \ifblank\subsectionnumstyle\then\else\numstyle\subsectionnum.\fi
     \numstyle\tablenum
     \numstyle\subtablenum}}


\referencestyle{sequential}                      
\referencenumstyle{arabic}                       
\def\putrefformat{\numstyle\referencenum}        
\def\referencenumformat{\numstyle\referencenum.} 
\def\putreferenceformat{
     \everypar={\hangindent=1em \hangafter=1 }%
     \def\\{\hfil\break\null\hskip-1em \ignorespaces}%
     \leftskip=\refnumindent\parindent=0pt \interlinepenalty=1000 }


\normalsize


\def\fmtversion{2.6M (June 1992)}

\catcode`\@=12


\topmargin= 1truein\vsize=9truein
\leftmargin=1truein\hsize=6.5truein

\baselinestretch=1400

\foot={\hfil\normalfonts\numstyle\pagenum\hfil}
\head={\hfil}

\pagenum=0\pagenumstyle{arabic}

{\pagenumstyle{blank}

{\rightline{\hbox to 4.5cm{\vtop{\hsize= 4.5cm
\baselinestretch=960\footnotefonts
\hfill\\
OHSTPY-HEP-T-94-024\\
CALT-68-1878\\
DOE/ER/01545-618}}{\hbox to .35cm{\hfill}}}}

\vskip 1.5 truecm
\footnotenumstyle{symbols}
\footnotenum= 0

\centertext{{\bf String Cosmology and the Dimension of Spacetime}\footnote{Work
supported in part
by the U.S. Department of Energy under Contract no. DEAC-03-81ER40050.}}

\vskip 1.2 truecm

\centertext{{Gerald B.~Cleaver}\footnote{gcleaver@pacific.mps.ohio-state.edu
after 1 September 1993}}

\vskip .3 truecm
\centertext{{\it The Ohio State University, Columbus,} OH 43210}
\vskip .1 truecm
\centertext{and}
\vskip .1 truecm
\centertext{{\it California Institute of Technology, Pasadena,} CA, 91125}
\vskip .3truecm
\centertext{and}
\vskip .3truecm
\centertext{Philip J.~Rosenthal}
\vskip .3 truecm
\centertext{{\it Harvard University, Boston,} MA 02138}
\vskip .1 truecm
\centertext{and}
\vskip .1 truecm
\centertext{{\it California Institute of Technology, Pasadena,} CA, 91125}

\vskip 1.2 truecm
\centerline{\bf Abstract}
\vskip .3 truecm
The implications of string theory for understanding the dimension of
uncompactified spacetime are investigated. Using recent ideas in string
cosmology, a new model is proposed to explain why three spatial dimensions
grew large. Unlike the original work of Brandenberger and Vafa, this
paradigm uses the theory of random walks. A computer model is
developed to test the implications of this new approach. It is found that
a four-dimensional spacetime can be explained by the proper choice of
initial conditions.
\hfill\vfill\eject}

%
%
\def\h{\hbox to .5cm{\hfill}}
\def\hof{\hbox to .15cm{\hfill}}
\def\hi{\hbox to .2cm{\hfill}}
\def\htf{\hbox to .35cm{\hfill}}

\def\ha#1{\n\hbox to .6cm{\n {#1}\hfill}}
\def\hb#1{\indent\hbox to .7cm{\n {#1}\hfill}}
\def\hc#1{\indent\hbox to .7cm{\hfill}\hbox to 1.1cm{\n {#1}\hfill}}
\def\hd#1{\indent\hbox to 1.8cm{\hfill}\hbox to 1.4cm{\n {#1}\hfill}}


\def\mpr#1{\markup{[\putref{#1}]}}
\def\pr#1{[\putref{#1}]}
\def\pe#1{(\puteqn{#1})}
\def\n{\noindent}

\def\ref{\reference}
\def\referencenumformat{[\numstyle\referencenum]}
\def\subon{\subequationnumstyle{alphabetic}}
\def\suboff{\subequationnumstyle{blank}}
\def\undbib{\underbar {\hbox to 2cm{\hfill}}, }
%
%
\def\a{\alpha^{'}  }
\def\al{\alpha}

\def\dag{ \dagger }

\def\e{{\rm e}}
\def\ep{\epsilon}

\def\gsim{{\buildrel >\over \sim}}

\def\sqr{\sqrt}

\def\theta{\vartheta}

%
%
\def\half{{1\over 2}}

%
%

\def\BVT{{ Brandenberger, Vafa, and Tseytlin} }

\def\ie{{\it i.e.}}

%
%

\def\NPB{ {\it Nucl.~Phys.~}{\bf B}}


\font\twelveBbb=msym10 scaled \magstep1
\font\nineBbb=msym9
\font\sevenBbb=msym7
\newfam\Bbbfam
\textfont\Bbbfam=\twelveBbb
\scriptfont\Bbbfam=\nineBbb
\scriptscriptfont\Bbbfam=\sevenBbb
\def\Bbb{\fam\Bbbfam\twelveBbb}

 \def\R{{\Bbb R}} \def\S{{\Bbb S}} \def\T{{\Bbb T}}
   
 \def\Z{{\Bbb Z}}

\pagenum=0
\pagenumstyle{arabic}
\footnotenumstyle{arabic}
\footnotenum= 0

\sectionnumstyle{arabic}
\subsectionnumstyle{blank}
\sectionnum=0
\section{\bf Introduction}
\equationnum=0

In spite of extraordinary successes, traditional cosmology has left
unanswered a number of fundamental questions and been plagued by potential
inconsistencies. Arguably the most troubling problem is the pointlike
initial singularity at the time of the big bang, ``$t\equiv 0$''.
Almost equally distressing is the related prediction of infinite initial
temperature. Frequently, one sidesteps the preceding problems by appealing
to some
future theory of quantum gravity. A classical theory, general relativity is
expected to break down at small scales where quantum effects should
dominate. Thus, the divergences predicted as $t\rightarrow 0$ by the standard
Friedman-Robertson-Walker cosmology are expected to be artifacts of using
a classical theory in a quantum regime. One hopes that
as this limit is approached,
a proper theory of
gravity would predict a small but non-singular universe,
which would have no divergent physical quantities. Indeed, such
an outcome can be seen as a test for any candidate theory of quantum
gravity.

An equally compelling, albeit less common, open question in
traditional cosmology is why we live in a four-dimensional universe. While
many are content to insert the dimension of spacetime by hand, it would be
more satisfying to explain its value.

One need no longer talk about quantum gravity as a distant dream; with the
advent of string theory, we have a
candidate theory of quantum gravity today and therefore an
unrivaled potential tool for understanding cosmology. Conversely, cosmology
provides a unique arena for testing string theory's performance as a theory
of quantum gravity. Since string theory may make
qualitatively different predictions than point particle theories, one can
hope that some of the consequences are observable and will lead to the
first experimental (or at least observational) tests of string theory.

Indeed, as will be seen below, string theory  resolves the
problem of an initial pointlike singularity. Furthermore, it goes a long
way towards providing a maximum possible finite temperature for the universe.
Arguments are also being developed for why there are three ``large''
spatial dimensions in our universe, rather than nine or $25$, etc..
Finally, string theory may suggest solutions to many other cosmological
problems, as it naturally could provide ``cosmic'' strings, other sources of
dark matter and ultimately a resolution to the cosmological constant
problem.

\section{\bf Duality and the Initial Singularity}
\equationnum=0

String theory resolves the problem of an initial
singularity. Using the duality symmetry of string theory, we
see that the smallest possible radius of the universe has some non-zero
value. This minimum radius is the fixed-point of the duality transformation.
Duality is most easily demonstrated by considering the mode
expansion\mpr{schwarz87} of a compactified bosonic string
coordinate:
$$X=x+({m\over 2R}+nR)(\tau+\sigma)+({m\over 2R}-nR)(\tau-\sigma)+{\rm
oscillators}
\eqno\eqnlabel{coordexp}$$
where $m$, $n\in\Z$. (Note we have chosen $\sqrt{\a}= \sqrt{\half}$ in units of
the Planck length, $l_{\rm Pl}\equiv {\sqrt{\hbar G_{\rm
N}/c^3}}$.)\footnote{For those
who prefer that $\a$ remain a free parameter, the
expansion appears in the following more general form (with
oscillator modes also explicitly given):
$$\eqalign{
X(\sigma,\tau)
   &= x+ 2\a p\tau + 2L\sigma +
      {{i \over 2}{\sqr{2\a}}}\sum_{n\neq 0} {1\over n}
      (\alpha_n \e^{-2in(\tau+\sigma)}
      + \tilde{\alpha}_n \e^{-2in(\tau-\sigma)} )\cr
   &= X_{\rm r} + X_{\rm l},\,\, {\rm where}\cr
X_{\rm r}
   &= x_{\rm r} + {{2\a}}p_r(\tau-\sigma)
      + {i \over 2}{\sqr{2\a}} \sum_{n\neq 0}
      {1\over n}\, \alpha_n \e^{-2in(\tau-\sigma)}\,\, ,\cr
X_{\rm l}
   &= x_{\rm l }+ {{2\a}}p_l(\tau+\sigma)
      + {i \over 2}\sqr{2\a} \sum_{n\neq 0}
      {1\over n}\, \tilde{\alpha}_n \e^{-2in(\tau+\sigma)}\,\, .\cr}$$
Here $p={m\over R}$, $L=nR$,
$p_r= {1\over {{2}}}( p + {1\over \a }L)$, and
$p_l= {1\over {{2}}}( p - {1\over \a }L)$.}
We see that the left- and right-moving
momenta are,
$$(p_L,p_R)=({m\over 2R}+nR,{m\over 2R}-nR).\eqno\eqnlabel{momenta}$$
The first term, ${m\over 2R}$, is interpreted as one half the center of
mass momentum of the string, while the second term, $\pm nR$ is the
winding mode ``momentum.'' The corresponding string mass spectrum is,
$${1\over 4}M^2=N+{1\over 2}({m\over 2R}-nR)^2-1+\tilde N+{1\over 2}
({m\over 2R}+nR)^2-1.\eqno\eqnlabel{spectrum}$$
If we let $R\rightarrow {\a\over R}$, while simultaneously $m\leftrightarrow
n$, the spectrum is preserved. Indeed, the scattering amplitudes also
respect ``$R\leftrightarrow {\a\over R}$ duality,'' and it has been shown
that replacing $R$ with ${\a\over R}$ produces an isomorphic conformal field
theory.\mpr{kikkawa84,sakai86} By duality, a pointlike universe is
equivalent to one that is infinitely
big. Thus, the ``smallest'' universe possible has
$R=1$ in units of ${\sqr{\a}}\approx$ the Planck length,
which is of the same order of the minimum
size expected to result from quantum gravity.
(Unless otherwise noted, we express
$R$ in units of ${\sqr{\a}}$.)

\section{\bf The Original Paradigm of Brandenberger and Vafa}
\equationnum=0

The basic framework of string cosmology rests on a reversal of the usual
compactification scenario. Rather than begin with $D_{\rm crit}$ uncompactified
({\ie}, large) dimensions and posit the spontaneous compactification of
$D_{\rm crit}-4$ of them, Brandenberger and Vafa\mpr{vafa1}
adopt a view more compatible with the cosmological idea of a small
universe that expands. They assume that the universe began with all
$D_{\rm crit}-1$ of its spatial
dimensions compactified near the Planck radius. One then tries to explain why
precisely
three of the dimensions became very large (``decompactified'') in a stringy
big bang. Since the universe has not expanded infinitely since the big
bang, we anticipate that all of its spatial dimensions are still
compactified today. Some simply have a larger radius of compactification
than others. This view allows strings to have winding modes about all
spatial dimensions.

In their seminal paper,\mpr{vafa1}
Brandenberger and Vafa suggested a tantalizing scenario in which winding
modes, a purely stringy phenomena, could be used to explain the dimension
of spacetime. They argued heuristically
that winding modes exert a negative pressure on the universe, thereby
slowing and ultimately reversing the expansion.
Since winding mode energy is linear with the radius of compactification
({\it i.e.}, the
scale factor of the universe), there is a large energy cost to
expanding with winding modes present. The question becomes ``in how many
dimensions can
winding modes be expected to interact frequently enough to annihilate?'' If
the universe expands in a dimension where annihilation is incomplete, the
windings will eventually force a recollapse of the universe to and possibly
past $R=1$ (which by duality can be interpreted as another attempt at
expansion). Note that in many ways the model is incomplete because no
mechanism or even justification is given for expansion. One could imagine
that the universe sits at the Planck scale forever. This makes the model
very hard to test or constrain.

Furthermore, one might be troubled that
there are many examples of cosmological features, like the cosmological
constant and domain walls, which seem to require energy during expansion,
but on closer analysis actually promote expansion. Indeed, Einstein's
equations lead us to believe that all matter/energy
should increase the expansion rate by contributing to $\rho$.
Brandenberger and Vafa side-stepped the issue by proclaiming that
Einstein's equations are invalid because they do not respect duality.
Furthermore, it was implied that if ordinary matter accelerates expansion,
windings (i.e., ``dual matter'') should stop expansion.

Fortunately, in reference \pr{vafa2} Tseytlin and Vafa show more convincingly
that winding modes do indeed inhibit
expansion by studying the low energy expansion of the tree level
gravitational-dilaton effective action:
$$S_0=-\int d^Dx{\sqrt{-G}}e^{-2\phi}[R+4(\partial\phi)^2].
\eqno\eqnlabel{seffective}$$
Here, $D$ is the total dimension of spacetime. Now consider a time
dependent dilaton
and a metric of the form
$$ds^2=-dt^2+\sum^{D-1}_{i=1}a_i^2(t)dx_i^2\eqno\eqnlabel{metric}$$
in the presence of a string gas in thermal equilibrium at temperature
$T={1\over\beta}$. Defining $d=D-1$, $a_i(t)=e^{\lambda_i(t)}$ and
$\varphi=2\phi-\sum_{i=1}^d\lambda_i$, and truncating to zero modes,
the full action becomes,
$$-\int dt{\sqrt{-G_{00}}}[e^{-\varphi}(-G^{00}
\sum^d_{i=1}\dot\lambda_i^2+G^{00}\dot\varphi^2)-
F(\lambda_i,\,\beta{\sqrt{-G_{00}}})]\eqno\eqnlabel{fullaction}$$
where a term, $F$, has been added to include the free energy of matter.
The equations of motion are found by varying with respect to $G_{00}$,
$\lambda$ and $\varphi$. Rewritten in terms of the original dilaton $\phi$,
the equations of motion are:
\subon
$$\eqalignno
{-\sum_{i=1}^d\dot\lambda^2_i+(2\dot\phi-\sum_{i=1}^d\dot\lambda_i)^2
&=e^{2\phi}\rho &\eqnlabel{em-a}\cr
\ddot\lambda_j-(2\dot\phi-\sum_{i=1}^d\dot\lambda_i)\dot\lambda_j
&={1\over 2}e^{2\phi}p_j &\eqnlabel{em-b}\cr
2\ddot\phi-\sum_{i=1}^d\ddot\lambda_i-\sum_{i=1}^d\dot\lambda_i^2
&={1\over 2}e^{2\phi}\rho. &\eqnlabel{em-c}\cr}$$
\suboff
Above, $\rho={E\over V}$ and $p_i={P_i\over V}$ is the pressure where
$V=\exp(\sum_i\lambda_i)$, and $E$ and $P_i$ are defined in terms of the free
energy,
$$\eqalign{E &=F+\beta{\partial F\over \partial\beta}\,\, ,\cr
P_i &=-{\partial F\over \partial\lambda_i}\,\, .}\eqno\eqnlabel{ep}$$
In order to solve these equations, we specialize to the isotropic case
where all the $\lambda_i$ are taken to be equal and $P\equiv -{1\over
d}{\partial
E\over \partial\lambda}$. Then eqs.~(\puteqn{em-a}-c)
reduce to
\subon
$$\eqalignno
{-d\dot\lambda^2+\dot\varphi^2
&=e^{\varphi}E &\eqnlabel{em2-a}\cr
\ddot\lambda-\dot\varphi\dot\lambda
&={1\over 2}e^{\varphi}P &\eqnlabel{em2-b}\cr
\ddot\varphi-d\dot\lambda^2
&={1\over 2}e^{\varphi}E. &\eqnlabel{em2-c}\cr}$$
\suboff
These equations can now be solved once initial conditions are chosen, if
$E(\lambda)$ is known. Unfortunately, this function is not well understood.
However, a universe with much of its energy in windings will have $E\sim R$
so it is reasonable to assume that for relatively large $R$
$$E(\lambda)=e^{\alpha\lambda}\,\, ,\eqno\eqnlabel{eofr}$$
where $\alpha$ is of order unity.

The solution is easiest when $\alpha=0$, as is appropriate in the very
early oscillator dominated regime of the Brandenberger-Vafa-Tseytlin model.
Then the dilaton and radius vary by
\subon
$$\eqalignno{
e^{-\varphi} &={Et^2\over 4}-{dA^2\over E} &\eqnlabel{alphzero-a}\cr
\lambda &=\lambda_0+\ln\left({t-2{\sqrt{d}}A/E\over t+2{\sqrt{d}}A/E}\right)
&\eqnlabel{alphzero-b}}$$
\suboff
where $A$ is an integration constant. We see that even in the extreme
case of $\alpha=0$, the expansion slows and
ultimately is halted. When $\alpha>0$ it has been shown that not only is
the expansion stopped in finite time, but it is also reversed.
Thus we see that winding strings
must collide and annihilate for significant and continued expansion to
occur.

In what number of spacetime dimensions can annihilation be expected?
Clearly, annihilation is easier in fewer dimensions. For example, in $1+1$
dimensions, the windings must lie on top of each other. In $2+1$
dimensions, they can be separated by one coordinate, but would be expected
to interact frequently. In a very large number of
dimensions, one would expect the equilibrium between winding modes to most
likely be lost, so that their number density need not fall drastically as
their energy increases. What is the maximum spacetime dimension which would
allow thermal equilibrium between winding modes and thus lead to their total
annihilation during expansion? Many\mpr{vafa1,polchinski88,khuri92}
have argued that a pair of $2$-dimensional
world sheets should generically intersect in four or fewer spacetime
dimensions because $2+2=4$. Even if four were a rigorously proven maximum,
the question would remain why four dimensions are favored over a smaller
number, which seems much more likely from the point of view of ease of
interaction. It has been suggested that entropy considerations would favor four
dimensions.\mpr{cateau92} However, to our knowledge,
no attempt has yet been
made to demonstrate in detail how four spacetime dimensions are dynamically
favored for winding mode annihilation.

The following sections consider a new model for understanding origins of the
dimension of spacetime, that has its roots in the preceding discussion. The
next section argues, using the theory of random surfaces and walks, that
four is indeed the maximum dimension of large (decompactified) spacetime, if
the background
spacetime is quasi-static and Euclidean. Possible modifications for a more
realistic spacetime are also discussed. In the last section, a
computer model is described that was used to explore qualitatively the
implications of this new paradigm. The goal is to test the viability of the
model by
asking if it could yield sensible results. Detailed predictions must await
a fuller understanding of string interactions and string thermodynamics at
extreme temperatures.\footnote{For a review of work in this area, see the
forthcoming ref.~\pr{cleaver93d}.}
Nevertheless, from our model approximate limits on the magnitude of the
Hubble expansion are
found which are consistent with theoretical estimates.\mpr{wise84}
Furthermore, it is
found that the preferred dimension of spacetime need not be two, as might
have been inferred from the original model of Brandenberger and Vafa.

\section{\bf Dimensional Predictions and Random Walks of Winding Modes}
\equationnum=0

As discussed in section three, we probably do exist in a
large four-dimensional universe because $2+2=4$,
but due to more involved physics than the initial argument
applying this to strings implies.
The world sheets of strings with winding modes
(and string world sheets as a whole) are not simple planes,
especially at high energies. Instead they may
have many complex bends and twists, at least at high energies.
At or near the Planck scale it is probably more accurate to model strings
as random surfaces.  In point particle field theory,
correlation functions can be bounded by the intersection
properties of two random walks. Whether or not two random walks will
intersect in a $D$-dimensional embedding space.
depends of their Hausdorff dimension, $d_{\rm H}$.

Strictly,
the Hausdorff dimension of a surface is found by considering a
covering of the surface by boxes of size $\epsilon$ and defining,
$$l_d(\epsilon)=\inf\sum_i\epsilon_i^d\,\, ,\eqno\eqnlabel{ld}$$
where the infimum is over all choices of $\epsilon_i$ with
$\epsilon_i<\epsilon$.
The Hausdorff dimension is
then defined implicitly by
$$l_d=\lim_{\epsilon\to0}l_d(\epsilon)=
\cases{0, &for $d>d_{\rm H}$;\cr
       \infty, &for $d<d_{\rm H}$.}
\eqno\eqnlabel{dh}$$
The Hausdorff dimension is extremely difficult to determine. Thus, one
typically finds instead the capacity
(fractal) dimension, $d_c$, which is
defined by
$$d_c=\lim_{\epsilon\to0}{\ln N(\epsilon)\over
\ln({1\over\epsilon})}\,\, ,\eqno\eqnlabel{dc}$$
where $N(\epsilon)$ is the number of boxes of size $\epsilon$ needed to cover
the surface. In most cases the capacity dimension, $d_c$, and $d_{\rm H}$
are equal. Rigorously, what one knows is that $d_{\rm H}\le
d_c$.\mpr{barnsley88}
For our purposes, we use the more common
intuitive definition of $d_{\rm H}$ which is that
$<X^2>\sim N^{{2/ d_{\rm H}}}$ for a walk of $N$ steps.
This last definition will suffice for most occasions, and it agrees with
the more rigorous mathematical definitions in all but singular
cases.\mpr{distler90}
For random surfaces,
$<X^2>$ becomes the average square size of the object, the hypersurface,
that is formed by the walk and
$N$ is the number of faces of dimension $d_{\rm geo}$ forming the
hypersurface.\mpr{distler90}

Thus, the Hausdorff
dimension indicates how the size of the walk (surface) scales,
which relates to
how ``space filling'' the walk (surface) is. Clearly, the higher the Hausdorff
dimension, the slower the size grows and the better the embedding space is
filled. If two random surfaces of geometrical dimension $d_{\rm geo}$ and
Hausdorff
dimension $d_{\rm H}$ are moving in an embedding space $\R^D$,
there will be a non-trivial intersection of the two surfaces
if $2d_{\rm H}>D$.
If $2d_{\rm H}<D$ the probability of intersection is low.
In the latter case, the two random walks (surfaces)
will each fill their own $d_{\rm H}$
dimensional subspaces and not likely overlap with each other.
The remaining
possibility, is the boundary case, $2d_{\rm H}=D$. Here the outcome is not so
certain {\it a priori}.
In some sense, the Hausdorff dimension describes the geometry of the random
surfaces as viewed by the embedding space. For random walks of points, the
Hausdorff dimension is two,
independent of the embedding
dimension.\mpr{gross84}

Thus, we suggest string interactions can be studied by considering the
intersection properties of random surfaces.
Whether two free
strings will interact in $D$-dimensions depends on the sum of the Hausdorff
dimensions of two random surfaces.
Surprisingly, the Hausdorff dimension of
a random two-dimensional surface is infinite for $D\geq 1$.\footnote{There
is actually some disagreement as to the correct Hausdorff dimension
for two dimensional random surfaces.
For random surfaces embedded in zero- and one-dimensional space,
Distler {\it et al.}\mpr{distler90} have shown that the Hausdorff
dimension, $d_{\rm H}$, is related to the embedding dimension, $D$, by
$$d_{\rm H}= {24\over {\sqrt{(25-D)(1-D)}} - (D-1)}\,\, .$$
Above $D=1$ there is disagreement though.
Gross argues that the Hausdorff dimensions for random surfaces is
infinite for embedding dimension $D>1$;\mpr{gross84} whereas,
Monte Carlo experiments of random surfaces suggest
$d_{\rm H}$ ranges monotonically from eight to ten for
embeddings in two- to ten-dimensional
space.\mpr{kazakov85}
In any case, this  latter range for $d_{\rm H}$ would
still suggest that
strings should interact in at least up to 16 dimensions.}
That is,
$$<X>_{{\rm random}~d_{\rm geo}=2~{\rm surface}}\sim \ln \, N\,\, .$$
(Generally, infinite Hausdorff dimension, $d_{\rm H}=\infty$,
is reserved to mean
 $\ln \, N$, not $N^0$.)\mpr{gross84,distler90}
This is the slowest possible non-zero
scaling of size as a function of the number of
steps.
Naively, this means that two strings will
interact when embedded in any number of dimensions.
This would suggest that strings are as likely to interact in ten
dimensions as in four.

Is this result valid for winding strings?
For high energy non-winding strings with wild fluctuations and twists
in spacetime it is, indeed, reasonable.
However, for strings with  winding modes
around a dimension with growing radius,
the Hausdorff dimension should quickly reduce to two.
Winding strings will not move freely in all directions; they will lie
basically parallel to the direction they are wound about.
Granted, if winding strings also have high energy oscillations, they may
also bend far away from parallel. But
transverse fluctuations of a
winding string quickly die down, through emission of low energy gravitons,
resulting from self-intersection of the string.
We can understand this most simply from the qualitative argument
that, since the central difficulty for expansion is the energy cost of
expanding with windings present, we would expect the oscillation energy to
be minimized. Although a
typical non-winding string of length $l$ will have fluctuations of the order
$\sqrt{l}$, it can be proven
that a winding string of the same length, deprived of high energy
oscillations, will only have fluctuations of the order $\langle \omega
\rangle$, where
$$\langle \omega \rangle^2= c_0 + {1\over 4\pi \sigma}\ln\,(l^2/\a)\,\, .
\eqno\eqnlabel{interlength}$$
Here $\sigma$ is the string tension and $(c_0)^{\half}$ is a constant of order
$1\,l_{\rm Pl}$,\mpr{polchinski93}
and like $R$, $\langle \omega \rangle$ is expressed
in units of ${\sqr{\a}}$.
This allows us to treat winding modes not as random surfaces, but rather as
random walks in the $D-1$ spacetime dimensions orthogonal to the winding
direction. From the previous results for random walks, this would suggest
that interaction (self-cancellation) of winding modes is possible only if
$D-1<4$. This is quite in agreement with phenomenological results!

Be that as it may,
there are still problems with this approach.
First, one might suggest that a string with
winding modes can
rap around both
decompactifying dimensions and smaller non-decompactifying and intersect in
the latter subspace.
Second, random walks are generally done in static Euclidean embedding space,
quite unlike the early (stringy) universe.
The approach of \BVT has been to view the
early ten-dimensional spacetime universe as a nine-dimensional torus
tensored with time, {\it i.e.}
a topologically non-trivial $\T^9\times\R$.
Compactifying embedding space from $\R^{D-1}$ to $\T^{D-1}$ should
significantly enhance the
interaction rate and perhaps increase the dimension of embedding space
in which random walks will intersect. On the other hand, the early universe
underwent significant expansion, giving the opposite effect.
In section five we introduce a computer generated toy model of the early
universe that considers these factors and additional ones.
Following this, we use our model to
set bounds on the maximum expansion rate, $H_{\rm max}$, of the universe if
at least one
dimension is to completely decompactify and not be stopped by the presense
of winding modes.  We also use this model to determine the range of the
phase space of initial
conditions (primary starting radius $R_0$ and expansion rate $H$)
that give highest probability to three spatial dimensions decompactifying.

\section{\bf The Computer Simulation}
\equationnum=0

Treating all these effects analytically is prohibitively difficult, since we
are forced to consider the universe near or even at the Planck scale.
Without a well developed theory of quantum gravity, one may doubt the
plausibility of analytically proving that exactly three spatial dimensions
are expected to
expand. Additionally, a proper treatment of the creation rate of windings
requires a knowledge of string thermodynamics well beyond the current level
of understanding. The first difficulty is
that near the Hagedorn temperature, $T_{\rm H}$,
the microcanonical ensemble must be
used, which ostensibly requires counting all the states in the universe.
Progress has been made in limiting regimes, but general results for
independent radii of varying size have yet to be exhibited and are sure to be
unwieldy at best. Furthermore, the inclusion of strong gravitational effects,
appropriate
for the early universe, would lead one to question the validity
of using thermodynamics. Even if a careful thermodynamic treatment of
winding creation in an equilibrium ensemble were possible, it would leave
unanswered the most interesting question: how does the winding creation
drop as equilibrium is lost?
This would
require understanding non-equilibrium statistical mechanics in the early
universe. Another problem is that the stability of the very topology of
spacetime has come into question at extreme temperatures. Above the
Hagedorn temperature, the conservation of winding number cannot be
guaranteed.\mpr{atick88,witten92}

In spite of the preceding difficulties, it is feasible to test whether this
model
for the expansion of the universe can work. In other words, we can ask
whether we can make reasonable phenomenological assumptions about various
processes in the early universe which in this model would lead to a strong
prediction that three dimensions expand. This would not prove that the
paradigm of Brandenberger and Vafa does work, but that it can. Furthermore,
one can  turn the problem around, asking what must the early universe
have been like in order to produce our four-dimensional spacetime.
Ultimately, useful constraints may be placed on the expansion
rate, the radius at which equilibrium is effectively lost, the number of
windings
surviving at that radius, interaction rates, temperature and other
quantities in the early universe by this procedure.

In this spirit, a computer model was developed to simulate
winding string
collisions in the early universe. While one would like to follow the model
from $t=0$, it is expected to be much more reliable below $T_{\rm H}$, where we
can more confidently use a string description and assume that oscillations
are suppressed. Thus, we begin evolving the model soon after the
temperature has dropped somewhat below $T_{\rm H}$, in an inflationary era.
The primary
goal is to better understand the evolution of the universe just after
the equilibrium of winding strings is lost.

Based on the previous discussion, the
windings about each dimension are represented by points in $D-2$ spatial
dimensions, where $D$ is the total number of spacetime coordinates in the
theory,
including both those that stay small and those that become effectively
``decompactified.'' For the Type II superstring, $D=10$.
Likewise, one takes $D=10$ for the heterotic string,
assuming that internal degrees of freedom,
not extra compactified spatial coordinates, provide the extra central
charge for the left moving sector.
The $D=26$ bosonic case is not investigated, since we seek
a phenomenologically realistic model. Dimensions other than the critical
dimension are also considered to study the effect of
dimension on various processes. Since the radius
at which the temperature drops significantly below $T_{\rm H}$
and, hence, at which the windings drop out of equilibrium, is not well known,
the
appropriate starting radius for the model is not precisely known. Thus, the
initial radii of the spatial dimensions is left
variable, but is typically chosen to be a few Planck lengths. A large
initial radius would invariably lead to only one
decompactified spatial dimension.
The radii of compactification, which truly are quantum mechanical objects, can
be allowed to fluctuate, typically up to
$l_{\rm Pl}$ per time step.
Since our results indicate this effect is not very
significant, it is only incorporated into some of the trials. For simplicity,
the
fluctuations are taken to be independent of position.

A certain number of windings are presumed to remain in this epoch, but the
precise number is unknown, so the initial number of windings is also a free
parameter. Since the total number of high energy strings in
the early universe roughly equals the log of the energy,
the number of windings
about each direction should not be huge. If the primordial universe
contained precisely the energy in our observable universe, assuming
critical density, there would only
be $135$ energetic strings in the entire universe! Of course, this is an
extreme
lower bound. Nevertheless, since one does not expect all the strings in the
universe to be winding strings, the number of windings about each direction can
reasonably be assumed to be at most of order ten when equilibrium is lost.
Only windings of $\pm 1$ about a single direction are considered.
By the time temperatures below $T_{\rm H}$ are reached, we expect
strings with higher winding excitations about a given direction will have
decayed to strings with unit winding number, as
required by Boltzmann suppression. Perhaps more important is the
possibility of strings with single windings about more than one direction.
While these may have significance, they are not
tracked in this first modeling attempt, since although they may increase the
overall interaction rate and thereby may alter the quantitative results, they
should
not change the kinds of qualitative effects we seek to study. Indeed, to
create these strings and allow them to participate in annihilation
interactions requires two separate interactions with a correspondingly
reduced net
probability. While it is true that multiple winding strings would execute
walks in effectively fewer dimensions, the enhancement is not expected to be
extreme. Cosmic string studies indicate inclusion of
multiple winding strings in our model would only increase the
interaction rates by 20 to 30 percent.\mpr{polchinski93}

The net winding number is set to zero, in order to satisfy
observational constraints on the isotropy of the
universe.\mpr{turok87a,turok87b}
The windings then execute a random walk, stepping up to $l_{\rm Pl}$ in each
time step of $\delta t = 1$ Planck time unit $\equiv t_{\rm Pl}$.
During each step, the computer checks for
annihilations of pairs of winding modes with opposite winding number.
Unlike most work in this field that assumes an ideal gas, this analysis
explicitly allows interactions.
Naively, if two windings come within $l_{\rm Pl}$, they can be expected to
annihilate.\mpr{vafa1,polchinski88,polchinski93} However,
as the length of the string and thus its energy grow, oscillations cost
less and less energy, compared to the total energy of the string so that
the effective thickness of the string
increases. Owing to quantum correlations, interactions are expected for
windings that come within
$\langle\omega\rangle$ of each other.\mpr{polchinski93}
(See eq.~(\puteqn{interlength}).)
Note that having the extra oscillations in no way violates our assumption
regarding the
straightness of the winding strings, since the scale of the oscillations
grows slowly compared to the size of the string.

The probability of
interaction, given a collision, may also vary with radius.
Using equations (\puteqn{alphzero-a}-b) one can show that the
coupling remains fairly constant for small radii in the limit
of $\al=0$ and all the radii are equal. More precisely, then
$$g^2=e^{2\phi}=K\left(1-({R\over
R_{\rm max}})^{{\sqrt{D-1}}}\right)^2.\eqno\eqnlabel{dileq}$$
$R_{\rm max}$ is the radius at which expansion stops and $K$ is an unknown
constant. We see that for
$R\ll R_{\rm max}$ the coupling remains constant to lowest order. For larger
$R$,
the coupling $e^{2\phi}$ decreases with radius, ultimately dropping to
zero. This agrees
with other studies that conclude there is only trivial scattering in
the infinite radius limit.\mpr{khuri92} One can use eq.~\pe{dileq} to get
an indication of the
importance the variation of the dilaton, even in the current context where the
radii are all independent, if one replaces $({R\over
R_{\rm max}})^{{\sqrt{D-1}}}$ by $({V\over V_{\rm
max}})^{{{1\over\sqrt{D-1}}}}$.
 At $R=1$, the
probability of annihilation, given a collision, is taken to be one,
providing the normalization to the coupling constant. Computer runs
are conducted assuming either a constant dilaton or a coupling varying by
eq. \pe{dileq}.  In
any case, the decrease in the coupling is not expected to be significant,
at least in ten
dimensions, since the dramatic drop in the collision rate with increasing
radius will dominate over any effect of the dilaton.

The universe is also allowed to expand during each time step.
 The proper expansion equation
can be found using equations (\puteqn{em-a}-c), provided that
one knew how the energy of the matter varied with independent radii. The
assumptions made by \pr{vafa2} that all the radii are equal and that
$E\sim R^{\alpha}$ clearly do not hold here. Eq. (\puteqn{alphzero-b}) is
also inappropriate since it shows all radii tending to a fixed value. No
dimensions effectively ``decompactify.'' Even if $E(R_i)$ were well
known, one would be forced to solve a system of $19$ coupled differential
equations to get a rigorous result. However, in order to understand the
qualitative implications of the model, one only needs to use an expansion
equation that has the correct features. The Brandenberger-Vafa framework,
verified in special cases by \pr{vafa2}, requires that the windings slow
the expansion as the radius increases and can ultimately stop or reverse
it. These essential features are captured by the following procedure:
During each time step, each radius $R_i$ is rescaled by a function of the
number of windings, $n_i$,
about that dimension and the radius itself,
$$R_i (t+1)=R_i (t) (1+\epsilon (n_i (t),R_i (t))).\eqno\eqnlabel{expand2}$$
If $\epsilon$ were independent of time, in the limit of $\delta t\rightarrow 0$
(\puteqn{expand2}) would approach
exponential expansion with constant Hubble parameter, $H= \ep =\dot R/R$.

This is, indeed, the form predicted by some authors to result from string
driven
inflation.\mpr{turok87a,turok87b,gasperini91}
In light of recent discussions  of a possible phase transition at the Hagedorn
temperature,\mpr{atick88,axenides88,lizzi90,lizzi91,salomonson86,turok87a,turok87b}
we will assume that in the absence of windings, $\ep$ is constant, whereas
in the presence of windings, that $\ep$ decreases  linearly with
increasing radius and with increasing number of windings.
Note that in refs.~\pr{vafa1, vafa2}
exponential expansion was not possible. However,
in \pr{vafa1, vafa2}
the possibility of inflation around $T_{\rm H}$
resulting from a stringy phase transition from a false vacuum
was not considered.
Some suggest this phase transition corresponds,
as the universe cools below $T_{\rm H}$, to individual strings
``condensing'' out of a single string or out of a string ``soup''.
At or above $T_{\rm H}$, this single string (``soup'')
fills all of spacetime and carries all, or nearly all, the energy.
Thus, in our model we combine the
ideas of exponential inflation with the expansion hindering effects
of winding modes. However, we do not expect our results to alter
significantly if instead of using exponential expansion, we chose
power-law expansions. Our results indicate that if winding modes are to
annihilate, they must do so very early,
early enough that exponential expansion is still subluminal
is closely approximated by a (low) power expansion.

The following form satisfies our requirements for $\epsilon$:
$$\epsilon (n_i (t),R_i (t))=H(1-{n_i (t)R_i (t)\over
2R_{\rm max}})\,\, ,\eqno\eqnlabel{eps}$$
where $H$
is the maximum expansion rate, as well as the Hubble constant for
de Sitter inflation, and $R_{\rm max}$ is the radius at which two windings will
halt
the expansion. $R_{\rm max}$ appeared previously in eq.~\pe{dileq}.
Clearly, $R_{\rm max}$ must be less than the radius at which GUT
physics takes place, but is not otherwise well constrained.
The importance and reasonable ranges of both
parameters will be determined by studying how they effect the prediction
for the dimension of spacetime. In our program, we do take
the limit of $\delta t\rightarrow 0$ and
replace $1+\ep(t)$ with $\exp(\ep(t))$ in eq.~\pe{expand2}.

This prescription yields an expansion rate
that decreases to zero as $R_i$ increases along directions with
corresponding windings present, but results in constant
exponential expansion about any dimension for which all the
corresponding windings have been annihilated.
Dimensions do not recontract if their windings remain, instead
staying compactified at $R_i= 2R_{\rm max}/n_i$.
Though an expansion equation that allows
contraction could be constructed, it would not be useful in the model
(as we discuss shortly).
It is of course
possible in a given expansion attempt for  no dimensions to lose all
their windings. In that case, the dimensions
are expected to recontract and ultimately begin expansion again.
However, for inflation to occur a second time, the universe must have
recontracted back into the (topological?) phase at or above
the Hagedorn temperature. We cannot follow the
windings as they enter this phase. When the universe begins expansion
again, one could begin modeling below the Hagedorn temperature as before.
This would be essentially treated as an independent attempt at expansion.
Thus, the model should be seen as following the evolution of the universe
during its {\it final and only successful attempt\/} at expansion. The above
reasoning also implies that if some dimensions lose all their windings,
then these will not stay forever at the Planck scale. With some dimensions
large,
the temperature can no longer grow high enough to restore the
compactified space back into its original vacuum, so that the inflation of the
small dimensions cannot be repeated.

Causality raises some questions about how to implement the preceding
prescription. These difficulties result from trying to incorporate the
effects of a purely global concept like winding number into local physical
effects like expansion. The number of windings about a given direction is
globally defined, irrespective of position. However, the effect of those
windings on local physics cannot change everywhere instantaneously.
By causality, if a
winding is annihilated at some spacetime point, $X^{\mu}$, one would expect
the expansion rate far away to be unaffected initially.
Strictly, winding annihilations would lead to growing bubbles of spacetime,
which are expanding at a faster rate then the rest of space.
After a
number of annihilations, each spacetime point would have expanded a
different amount. This is very difficult to model. Instead, a retardation
is introduced into our model so that $n_i$ in eq.~\pe{eps}
 counts windings that either have
not annihilated or have annihilated too recently for most of the universe
to know about it. Specifically, an annihilation is ``counted'' after a time
equal to the effective radius of the universe,
$R={\sqrt{\sum_{i=1}^dR_i^2}}$, at the time of the annihilation. Runs are
conducted both with and without the retardation to determine its
importance.

The very early universe should be hot enough to create pairs of winding
strings. However, it is highly nontrivial to compute that
rate. Following ref.~\pr{bowick92}, one could find both the
number of winding states and the total number of states for a gas of
strings in the high energy limit.
Ideally, one would want to relax the
equilibrium assumption to get more accurate results in the more interesting
era when equilibrium is not present. Indeed, having an equilibrium
description obviates the need for computer modeling, telling us exactly
how many windings should exist at any given time. A naive argument can show
that creation of windings must cease at a very small radius. If the
expansion process is roughly adiabatic, then $T\sim{1\over R}$. Furthermore,
the
energy of windings is linear with $R$. Thus, the Boltzmann supression
factor would go
as $e^{-R^2}$. Even though deviations from adiabaticity may occur, the
suppression is strong enough that one can believe the creation rate is
negligible in the relevant regime. Thus, we assume
creation of winding modes has ceased by the time the winding modes
have effectively fallen out of equilibrium and we begin our numerical trails.

\section{\bf Results of the Simulation and Predictions of the Model}
\equationnum=0

The central result of the computer simulation is that a two-dimensional
decompactified
universe need not be the most probable outcome of the model just presented.
If one considers the
full parameter space described in the last section, the vast majority of it
corresponds to either a two- or a ten-dimensional spacetime. However,
appropriate choices of parameters can be found to make any dimension, from
two to ten, the most probable. Unfortunately, in cases when the most likely
dimension is neither two nor ten, it generally
becomes impossible to predict the outcome with reasonable certainty.

Two dimensions result when annihilations are extremely unlikely. All
dimensions then have $n_i$ windings about them that survive so that each
dimension can at most expand to $2R_{\rm max}/n_i$. The simulation would then
show a
result of zero large spatial dimensions or a one-dimensional spacetime.
However, we know that given
sufficient time, annihilation must ultimately occur, since the space is no
longer rapidly expanding.
This time may have to be integrated over several
expansion attempts if it is more likely that the universe will
recollapse before such annihilations occur!
(See figure 2.)
In this case the entire
scenario would be repeated, presumably with the same choice of initial
parameters, since they are determined by the poorly understood physics of
the Planck scale. Once the rare annihilation occurs that leaves a dimension
without windings, this dimension will expand without bound, forever
suppressing annihilations along other dimensions. More than one large
dimension would require two rare annihilations to occur almost
simultaneously. If the annihilations along different dimensions occurred
at significantly different times, then the large spatial dimensions we
observe today would have undergone vastly different amounts of expansion. This
is
probably ruled out by the isotropy of our universe.

A ten dimensional spacetime is achieved when the parameters are such that
annihilation is extremely efficient once equilibrium is lost. Then all
winding strings are destroyed almost immediately and all dimensions expand
without constraint.

The more interesting situation occurs for a relatively narrow band of
parameter space in which winding annihilation is moderately likely.
(See figures 4 and 5.)
The most
important variable is the radius at which equilibrium is lost and the
simulation begins. The importance of radius can be seen by examining how
the collision rate falls with the radius of compactification  in various
dimensions. For walks in one spatial
dimension, one would expect the number of steps required for collision to
scale as the square of the radius of compactification. This generally
holds, especially at large radii. Deviations result from the logarithmic
growth of the size of the string with radius. At small radii, ${{\sqrt{\ln
R}}\over R}$ is not negligible, accounting for the greater deviation in small
spaces. Figure 2 shows how the number of steps required for a
pair of windings to annihilate (in just three dimensional compactified space)
with at least $98$ percent probability
varies with the radius of compactification. As expected, in more dimensions
the collision rate drops dramatically. In nine dimensions, roughly $180,000$
steps ({\it e.g.}, $180,000$ Planck time units)
are required to get a collision with $98$ percent probability at a constant
radius of only three. As a result,
we see that the winding creation rate must drop effectively to zero at a
radius not much larger than $R=1$, or the expansion rate must be small enough
so that
hundreds of thousands of time steps lead to negligible expansion. (An
expansion rate of $10^{-4}$
would increase the radius by a factor of
$6\times 10^7$ in $180,000$ time steps.) If not, windings would be created at a
radius where they
had little chance of annihilating, leading to a two dimensional universe.

This raises the question about how fast the universe can expand
without preventing winding collision and annihilation. To answer this
question, trials were conducted with the expansion rate taken to be a
constant, independent of the number of windings present. We then checked
to see how large the expansion rate could be such that two windings would
collide with $98$ percent probability before the radii of compactification
were clearly too large for annihilation to occur ($R>500$.) In three
spatial dimensions, a very large Hubble parameter (of order one) is allowed
if equilibrium  is lost at the rather improbable value, $R=1$. However, if
the proper initial radius for the model is $R=4$ then the maximum Hubble
parameter is about $10^{-4}$ in Planck units. (See figure 3.) With nine spatial
dimensions, as is appropriate for the superstring, the largest possible
Hubble parameter is around $10^{-5}$ for $R$ starting slightly above
1.
These constraints are not precise limits,
since the actual expansion rate is not constant, as assumed above, but gets
reduced
in the presence of windings. Thus, the maximum expansion rate without
windings could be larger. More complicated string processes than those
treated here could also increase the annihilation rate and allow greater
expansion. Nevertheless, the preceding analysis indicates that the expected
magnitude of the maximum expansion is very small.

Are the above values reasonable? Abbott and Wise\mpr{wise84}
have shown that the size of the
Hubble parameter is bounded by observations of the cosmic microwave background
radiation. Large scale anisotropy induced by gravitational waves is
dominated by physical wavenumbers of order the present Hubble constant,
$H_{\rm O}$. The amplitude of these waves in {\it any} generalized inflationary
cosmology is of order $H_{\rm HC}/m_{\rm Pl}$, where $H_{\rm HC}$ is the value
of the Hubble constant at the time when a wave of present physical
wavenumber $H_{\rm O}$ crossed the Horizon during the inflationary era.
These waves produce a $\delta\, T/T$ of the order of their amplitude,
$H_{\rm HC}/m_{\rm Pl}$.
Thus, a Hubble constant during
inflation greater than around $10^{-5}$ would allow gravitational waves of
sufficient amplitude to produce a microwave anisotropy greater than that
observed by COBE. It is pleasing that the model described here produces similar
bounds.

Of course, the most direct and revealing way to determine the effect of the
radius of compactification and Hubble parameter on the expected dimension
of spacetime is to simply run many ($50$) trials for various values of
these parameters and compute the average dimension obtained. Typically,
we find $\langle D\rangle=10$ up to some $R_1$ and then falls rapidly as a
function of $R$ up
until $R_2$, beyond which the expected dimension is two. Unless otherwise
specified, all following runs use $10$ windings about each
of nine compactified directions, an $R_{\rm max}$
(the maximum radius obtainable with two windings present) of $50$, and an
effective string width $\langle\omega\rangle$
chosen to equal $2\pi$ at $R=1$. If the Hubble parameter, $H$, is between $.1$
and $.01$, $R_1$ is equal to one and $R_2$ is a very small $1.5$.
Four-dimensional spacetime is then
the most probable only in the narrow range of
of $1.18 \leq R_0 \leq 1.21$ for $H=.1$ and of $1.22\leq R_0\leq 1.24$ for
$H=.01$.
Since it is
hard to believe that the temperature could have dropped sufficiently below
the Hagedorn temperature for the windings to have fallen out of equilibrium
so near to the dual radius, we again conclude that a small expansion
rate is necessary. For $H=.001$, the interesting range for $R_0$ has only
increases to between $1.2$ and $1.6$,
with four-dimensions most-probable between
$1.51$ and $1.58$.
If $H=10^{-4}$, $R_1$ and $R_2$ are $1.5$ and
$2.5$ respectively. The range for four dimensions is now
$1.99$ to $2.10$. We estimate the uncertainties on $R_{1,2}$ to
be about $.05$. For $H=10^{-5}$, $R_1$ and $R_2$ are extrapolated to
be $2.0\pm .1$ and $3.3\pm .1$, respectively, with
four dimensions most probable in the range of 2.8 to
2.93.\footnote{At the time of the release of this preprint
our computer
program is being modified to run on a
Cray Y-MP to reduce the run time $H=10^{-5}$ would require.}
Thus, for the dimension of spacetime considered solely as
a function of $H$ and $R_0$, only a very narrow
range of the parameter space predicts ``four'' as the outcome.

Another parameter upon which the final dimension of spacetime sensitively
depends is the effective width of a string, determined by $c_0$ in the
expression above eq.~\pe{dileq}.
(See figure 6.)
Even with $R=1$ and $H=.01$, we find
that $\langle D\rangle =1$ ({\it i.e.}, that only time is uncompactified)
for $c_0$ ranging from zero to $15$.
The expected dimension rises
rapidly as $c_0$ increases from $15$ to $30$. Of course, a
wider string should act equivalently to a narrower string in a smaller
space, so this behavior is not surprising.
Clearly, if a change in phase occurs as $R$ approaches 1,
then the interaction width of strings should be large at
$R=1$, especially
if this change in phase corresponds to all strings merging
into a single string filling all of compactified space.
Thus, unless otherwise specified, we select $c= 37.64$, giving
$\langle\omega\rangle = 2\pi\, R$ at $R=1$.

The number of windings surviving when equilibrium is lost has a variable
effect. If the initial radius is small, it has almost no effect. For
example, with $R=1.4$, and $H=.1$, $500$ trials were conducted with either
$2$, $10$, $50$ or $100$ windings about each dimension. Even with a
sensitivity of $.07$ in the average dimension, no statistically significant
change in the average
dimension was observed when the number of windings ranged from $10$ to
$100$.
We can argue that since the initial volume of the transverse
space was only about $15$, ten or more windings completely filled the
space. This implies that for sufficiently small radii
the total number is irrelevant
and, further, might lead us to
expect that almost all of the windings would annihilate, as they are guaranteed
to
be in close proximity.
In reality, most of the time no dimensions got large,
This is because adjacent windings often do not have opposite winding
number.
For larger initial radii, the effect of the number of windings is, however,
very
significant. When the initial radius is two and $H=.0001$, the average
dimension of spacetime drops by over four when the number of windings
increases from $10$ to $25$.

Other parameters are less significant. The radius at which two windings
stop expansion, $R_{\rm max}$, does not greatly affect the results. In many
cases,
varying $R_{\rm max}$ from $5$ to $100$ has no effect, above error. If the
initial
radius is close enough to $R_{\rm max}$, then this parameter can reduce the
expected dimension of spacetime by about one. (See figure 7.)
This is to be
expected since a larger $R_{\rm max}$ allows faster expansion for a
given number of
windings, resulting in less collisions and a smaller dimension. The effect
of the evolution of the dilaton was also considered. While it can sometimes
drop the expected dimension by one or two sigma, the effect is
insignificant compared to other uncertainties, so many trials are conducted
with a constant dilaton. This result gives us confidence that deviations
from the approximate dilaton evolution equation being used \pe{dileq},
will not significantly affect the results. The effect of radii
fluctuations was also studied.
When radii fluctuations were allowed, they had no effect
whatsoever if $R_i \gsim 1.3$,
so fluctuations were subsequently ignored for most runs.

Finally, the importance of a time delay to enforce causality was
determined. In almost all cases
the time delay had almost no
statistically significant impact on the results. For some trials
the time delay reduced the expected dimension by up to two sigma
(i.e., by $.4$) for $\langle D \rangle$ near five.
The minimal effect of the time delay
indicates that one need not be concerned with constructing a more realistic
time delay algorithm.

The average dimension of spacetime is by no means the only quantity that
should be studied. The width of the expected distribution of dimensions is
also critically important. When the average dimension is one (ultimately
two) or ten, the width can be arbitrarily small. However, for intermediate
values, $\sigma$ is roughly $1-1.5$. (See figure 5.) Thus, while it is
possible to have an average dimension of spacetime of four, one cannot rule
out other alternatives based on the gross initial conditions
of the universe. This lack of determinism is not pleasing.

The above analysis shows that there are a number of parameters that can be
tuned to produce any desired average dimension of spacetime. The maximum
expansion rate, the radius at which equilibrium is lost, the number of
strings remaining at this radius and the effective width of those strings are
certainly the most important. Unfortunately, a firm prediction for the
most probable dimension of spacetime is not possible from this model
because of the number of free parameters and the omission of possibly
important physical effects. Nevertheless, this work does demonstrate how
string theory can be used to make such a prediction. A more complete model,
properly incorporating as yet poorly understood physics, is clearly called
for. Lastly, the the narrow range of parameters that give a four-dimensional
universe should be seen as a blessing in disguise, rather than
a fine tuning disaster. Once our knowledge of some of the relevant
parameters improves, we can use the fact that we live in a four-dimensional
world in analysis as done above to determine the values of the remaining
parameters to good accuracy.

\sectionnumstyle{blank}
\n {\bf\section{Acknowledgements:}}

G.C. and P.R. offer their sincere thanks to (in alphabetical order)
Mark Bowick, Jaques Distler, Lance Dixon, Ramzi Khuri,
Joe Polchinski, Chris Pope, and Cumrum Vafa for helpful and informative
discussions.

\hfill\vfill\eject

\sectionnumstyle{blank}
\n {\bf\section{References:}}

\begin{putreferences}

\ref{abbott84}{L.F.~Abbott and M.B.~Wise, {\it Nucl.~Phys.~}{\bf B244}
(1984) 541.}

\ref{alvarez86}{L.~Alvarez-Gaum\' e, G.~Moore, and C.~Vafa,
{\it Comm.~Math.~Phys.~}{\bf 106} (1986) 1.}

\ref{antoniadis86} {I.~Antoniadis and C.~Bachas, {\it Nucl.~Phys.~}{\bf B278}
 (1986) 343;\\
M.~Hama, M.~Sawamura, and H.~Suzuki, RUP-92-1.}
\ref{li88} {K.~Li and N.~Warner, {\it Phys.~Lett.~}{\bf B211} (1988)
101;\\
A.~Bilal, {\it Phys.~Lett.~}{\bf B226} (1989) 272;\\
G.~Delius, preprint ITP-SB-89-12.}
\ref{antoniadis87}{I.~Antoniadis, C.~Bachas, and C.~Kounnas,
{\it Nucl.~Phys.~}{\bf B289} (1987) 87.}
\ref{antoniadis87b}{I.~Antoniadis, J.~Ellis, J.~Hagelin, and D.V.~Nanopoulos,
{\it Phys.~Lett.~}{\bf B149} (1987) 231.}
\ref{antoniadis88}{I.~Antoniadis and C.~Bachas, {\it Nucl.~Phys.~}{\bf B298}
(1988) 586.}

\ref{ardalan74}{F.~Ardalan and F.~Mansouri, {\it Phys.~Rev.~}{\bf D9} (1974)
3341; {\it Phys.~Rev.~Lett.~}{\bf 56} (1986) 2456;
{\it Phys.~Lett.~}{\bf B176} (1986) 99.}

\ref{argyres91a}{P.~Argyres, A.~LeClair, and S.-H.~Tye,
{\it Phys.~Lett.~}{\bf B235} (1991).}
\ref{argyres91b}{P.~Argyres and S.~-H.~Tye, {\it Phys.~Rev.~Lett.~}{\bf 67}
(1991) 3339.}
\ref{argyres91c}{P.~Argyres, J.~Grochocinski, and S.-H.~Tye, preprint
CLNS 91/1126.}
\ref{argyres91d}{P.~Argyres, K.~Dienes and S.-H.~Tye, preprints CLNS 91/1113;
McGill-91-37.}
\ref{argyres91e} {P.~Argyres, E.~Lyman, and S.-H.~Tye
preprint CLNS 91/1121.}
\ref{argyres91f}{P.~Argyres, J.~Grochocinski, and S.-H.~Tye,
{\it Nucl.~Phys.~}{\bf B367} (1991) 217.}
\ref{dienes92a}{K.~Dienes, Private communications.}
\ref{dienes92b}{K.~Dienes and S.~-H.~Tye, {\it Nucl.~Phys.~}{\bf B376} (1992)
297.}

\ref{athanasiu88}{G.~Athanasiu and J.~Atick, preprint IASSNS/HEP-88/46.}

\ref{atick88}{J.~Atick and E.~Witten, {\it Nucl.~Phys.~}{\bf B310}
(1988) 291.}

\ref{axenides88}{M.~Axenides, S.~Ellis, and C.~Kounnas,
{\it Phys.~Rev.~}{\bf D37} (1988) 2964.}

\ref{bailin92}{D.~Bailin and A.~Love, {\it Phys.~Lett.} {\bf B292}
(1992) 315.}

\ref{barnsley88}{M.~Barnsley, {\underbar{Fractals Everywhere}} (Academic
Press, Boston, 1988).}

\ref{bouwknegt87}{P.~Bouwknegt and W.~Nahm,
{\it Phys.~Lett.~}{\bf B184} (1987) 359;\\
F.~Bais and P.~Bouwknegt, {\it Nucl.~Phys.~}{\bf B279} (1987) 561;\\
P.~Bouwknegt, Ph.D.~Thesis.}

\ref{bowick89}{M.~Bowick and S.~Giddings, {\it Nucl.~Phys.~}{\bf B325}
(1989) 631.}
\ref{bowick92}{M.~Bowick, in
{\underbar{Strings, Quantum Gravity, and Physics at the Planck Energy
Scale,}}
N. Sanchez, ed. (World Scientific, Singapore, 1993) 44.}
\ref{bowick93}{M.~Bowick, Private communications.}

\ref{brustein92}{R.~Brustein and P.~Steinhardt, preprint UPR-541T.}

\ref{cappelli87} {A.~Cappelli, C.~Itzykson, and
J.~Zuber, {\it Nucl.~Phys.~}{\bf B280 [FS 18]} (1987) 445;
{\it Commun.~Math.~Phys.~}113 (1987) 1.}

\ref{carlitz72}{R.~Carlitz, {\it Phys.~Rev.~}{\bf D5} (1972) 3231.}

\ref{candelas85}{P.~Candelas, G.~Horowitz, A.~Strominger, and E.~Witten,
{\it Nucl.~Phys.~}{\bf B258} (1985) 46.}

\ref{cateau92}{H.~Cateau and K.~Sumiyoshi,
{\it Phys.~Rev.~}{\bf D46} (1992) 2366.}

\ref{christe87}{P.~Christe, {\it Phys.~Lett.~}{\bf B188} (1987) 219;
{\it Phys.~Lett.~}{\bf B198} (1987) 215; Ph.D.~thesis (1986).}

\ref{clavelli90}{L.~Clavelli {\it et al.}, {\it Int.~J.~Mod.~Phys.~}{\bf A5}
(1990) 175.}

\ref{cleaver92a}{G.~Cleaver. {\it ``Comments on Fractional Superstrings,''}
in the Proceedings of the International Workshop on String
Theory, Quantum Gravity and the Unification of Fundamental Interactions,
M. Bianchi {\it et. al}, ed. (World Scientific, Rome, 1992) 110.}
\ref{cleaver93a}{G.~Cleaver and D.~Lewellen, {\it Phys.~Rev.~}{\bf B300}
(1993) 354.}
\ref{cleaver93b}{G.~Cleaver and P.~Rosenthal, preprint CALT 68/1756.}
\ref{cleaver93c}{G.~Cleaver and P.~Rosenthal, preprint CALT 68/18__.}
\ref{cleaver}{G.~Cleaver, Unpublished research.}
\ref{cleaver93d}{G.~Cleaver. {\it String Cosmology: A Review},
To appear.}

\ref{cornwell89}{J.~F.~Cornwell, {\underbar{Group Theory in Physics}},
{\bf Vol. III}, (Academic Press, London, 1989).}

\ref{deo89a}{N.~Deo, S.~Jain, and C.~Tan, {\it Phys.~Lett.~}{\bf
B220} (1989) 125.}
\ref{deo89b}{N.~Deo, S.~Jain, and C.~Tan, {\it Phys.~Rev.~}{\bf D40}
(1989) 2626.}
\ref{deo92}{N.~Deo, S.~Jain, and C.~Tan, {\it Phys.~Rev.~}{\bf D45}
(1992) 3641.}
\ref{deo90a}{N.~Deo, S.~Jain, and C.-I.~Tan,
in {\underbar{Modern Quantum Field Theory}},
(World Scientific, Bombay, S.~Das {\it et al.} editors, 1990).}

\ref{distler90}{J.~Distler, Z.~Hlousek, and H.~Kawai,
{\it Int.~Jour.~Mod.~Phys.~}{\bf A5} (1990) 1093.}
\ref{distler93}{J.~Distler, private communication.}

\ref{dixon85}{L.~Dixon, J.~Harvey, C.~Vafa and E.~Witten,
 {\it Nucl.~Phys.~}{\bf B261} (1985) 651; {\bf B274} (1986) 285.}
\ref{dixon87}{L.~Dixon, V.~Kaplunovsky, and C.~Vafa,
{\it Nucl.~Phys.~}{\bf B294} (1987) 443.}

\ref{drees90}{W.~Drees, {\underbar{Beyond the Big Bang},}
(Open Court, La Salle, 1990).}

\ref{dreiner89a}{H.~Dreiner, J.~Lopez, D.V.~Nanopoulos, and
D.~Reiss, preprints MAD/TH/89-2; CTP-TAMU-06/89.}
\ref{dreiner89b}{H.~Dreiner, J.~Lopez, D.V.~Nanopoulos, and
D.~Reiss, {\it Phys.~Lett.~}{\bf B216} (1989) 283.}

\ref{ellis90}{J.~Ellis, J.~Lopez, and D.V.~Nanopoulos,
{\it Phys.~Lett.~}{\bf B245} (1990) 375.}

\ref{fernandez92}{R.~Fern\' andez, J.~Fr\" ohlich, and A.~Sokal,
{\underbar{Random Walks, Critical Phenomena, and Triviality in}}
{\underbar{Quantum Mechanics}}, (Springer-Verlag, 1992).}

\ref{font90}{A.~Font, L.~Ib\'a\~ nez, and F.~Quevedo,
{\it Nucl.~Phys.~}{\bf B345} (1990) 389.}

\ref{frampton88}{P.~Frampton and M.~Ubriaco, {\bf D38} (1988) 1341.}

\ref{francesco87}{P.~di Francesco, H.~Saleur, and J.B.~Zuber,
{\it Nucl.~Phys.~} {\bf B28 [FS19]} (1987) 454.}

\ref{frautschi71}{S.~Frautschi, {\it Phys.~Rev.~}{\bf D3} (1971) 2821.}

\ref{frohlich92}{R.~Fern\'andez, J.~Fr\"ohlich and A.~Sokal,
\underbar{Random Walks, Critical Phenomena, and}\\
\underbar{Triviality in
Quantum Field Theory}, (Springer-Verlag, Berlin, 1992).}

\ref{gannon92}{T.~Gannon, Carleton preprint 92-0407.}

\ref{gasperini91}{M.~Gasperini, N.~S\'anchez, and G.~Veneziano,
{\it Int.~Jour.~Mod.~Phys.~}{\bf A6} (1991) 3853;
{\it Nucl.~Phys.~}{\bf B364} (1991) 365.}

\ref{gepner87}{D.~Gepner and Z.~Qiu, {\it Nucl.~Phys.~}{\bf B285} (1987)
423.}
\ref{gepner87b}{D.~Gepner, {\it Phys.~Lett.~}{\bf B199} (1987) 380.}
\ref{gepner88a}{D.~Gepner, {\it Nucl.~Phys.~}{\bf B296} (1988) 757.}

\ref{ginsparg88}{P.~Ginsparg, {\it Nucl.~Phys.~}{\bf B295 [FS211]}
(1988) 153.}
\ref{ginsparg89}{P.~Ginsparg, in \underbar{Fields, Strings and Critical
Phenomena}, (Elsevier Science Publishers, E.~Br\' ezin and
J.~Zinn-Justin editors, 1989).}

\ref{gross84}{D.~Gross, {\it Phys.~Lett.~}{\bf B138} (1984) 185.}

\ref{green53} {H.~S.~Green, {\it Phys.~Rev.~}{\bf 90} (1953) 270.}

\ref{hagedorn68}{R.~Hagedorn, {\it Nuovo Cim.~}{\bf A56} (1968) 1027.}

\ref{kac80}{V.~Ka\v c, {\it Adv.~Math.~}{\bf 35} (1980) 264;\\
V.~Ka\v c and D.~Peterson, {\it Bull.~AMS} {\bf 3} (1980) 1057;
{\it Adv.~Math.~}{\bf 53} (1984) 125.}
\ref{kac83}{V.~Ka\v c, {\underbar{Infinite Dimensional Lie Algebras}},
(Birkh\" auser, Boston, 1983);\\
V.~Ka\v c editor, {\underbar{Infinite Dimensional Lie Algebras and Groups}},
(World Scientific, Singapore, 1989).}

\ref{kaku91}{M.~Kaku, \underbar{Strings, Conformal Fields and Topology},
(Springer-Verlag, New York, 1991).}

\ref{kawai87a} {H.~Kawai, D.~ Lewellen, and S.-H.~Tye,
{\it Nucl.~Phys.~}{\bf B288} (1987) 1.}
\ref{kawai87b} {H.~Kawai, D.~Lewellen, J.A.~Schwartz,
and S.-H.~Tye, {\it Nucl.~Phys.~}{\bf B299} (1988) 431.}

\ref{kazakov85}{V.~Kazakov, I.~Kostov, and A.~Migdal,
{\it Phys.~Lett.~}{\bf B157} (1985) 295.}

\ref{khuri92}{R.~Khuri, CTP/TAMU-80/1992; CTP/TAMU-10/1993.}

\ref{kikkawa84}{K.~Kikkawa and M.~Yamasaki, {\it Phys.~Lett.~}{\bf B149}
(1984) 357.}

\ref{kiritsis88}{E.B.~Kiritsis, {\it Phys.~Lett.~}{\bf B217} (1988) 427.}

\ref{langacker92}{P.~Langacker, preprint UPR-0512-T (1992).}

\ref{leblanc88}{Y.~Leblanc, {\it Phys.~Rev.}{\bf D38} (1988) 38.}

\ref{lewellen87}{H.~Kawai, D.~Lewellen, and S.-H.`Tye,
{\it Nucl.~Phys.~}{\bf B288} (1987) 1.}
\ref{lewellen}{D.~C.~Lewellen, {\it Nucl.~Phys.~}{\bf B337} (1990) 61.}

\ref{lizzi90}{F.~Lizzi and I.~Senda, {\it Phys.~Lett.~}{\bf B244}
(1990) 27.}
\ref{lizzi91}{F.~Lizzi and I.~Senda, {\it Nucl.~Phys.~}{\bf B359}
(1991) 441.}

\ref{lust89}{D.~L\" ust and S.~Theisen,
{\underbar{Lectures on String Theory,}} (Springer-Verlag, Berlin, 1989).}

\ref{maggiore93}{M.~Maggiore, preprint IFUP-TH 3/93.}

\ref{mansouri87} {F.~Mansouri and X.~Wu, {\it Mod.~Phys.~Lett.~}{\bf A2}
(1987) 215; {\it Phys.~Lett.~}{\bf B203} (1988) 417;
{\it J.~Math.~Phys.~}{\bf 30} (1989) 892;\\
A. Bhattacharyya {\it et al.,} {\it Mod.~Phys.~Lett.~}{\bf A4} (1989)
1121; {\it Phys.~Lett.~}{\bf B224} (1989) 384.}

\ref{narain86} {K.~S.~Narain, {\it Phys.~Lett.~}{\bf B169} (1986) 41.}
\ref{narain87} {K.~S.~Narain, M.H.~Sarmadi, and C.~Vafa,
{\it Nucl.~Phys.~}{\bf B288} (1987) 551.}

\ref{obrien87}{K.~O'Brien and C.~Tan, {\it Phys.~Rev.~}{\bf D36} (1987)
1184.}

\ref{parisi79}{G.~Parisi, {\it Phys.~Lett.~}{\bf B81} (1979) 357.}

\ref{polchinski88}{J.~Polchinski, {\it Phys.~Lett.~}{\bf B209} (1988)
252.}
\ref{polchinski93}{J.~Polchinski, Private communications.}

\ref{pope92}{C.~Pope, preprint CTP TAMU-30/92  (1992).}

\ref{raiten91}{E.~Raiten, Thesis, (1991).}

\ref{roberts92}{P.~Roberts and H.~Terao, {\it Int.~J.~Mod.~Phys.~}{\bf A7}
(1992) 2207;\\
P.~Roberts, {\it Phys.~Lett.~}{\bf B244} (1990) 429.}

\ref{sakai86}{N.~Sakaii and I.~Senda, {\it Prog.~Theo.~Phys.~}
{\bf 75} (1986) 692.}

\ref{salomonson86}{P.~Salomonson and B.-S.~Skagerstam, {\it
Nucl.~Phys.~}{\bf B268} (1986) 349.}

\ref{schellekens89} {A.~N.~Schellekens and S.~Yankielowicz,
{\it Nucl.~Phys.~}{\bf B327} (1989) 3;\\
A.~N.~Schellekens, {\it Phys.~Lett.~}{\bf 244} (1990) 255;\\
B.~Gato-Rivera and A.~N.~Schellekens, {\it Nucl.~Phys.~}{\bf B353} (1991)
519; {\it Commun.~Math.}
{\it Phys.~}145 (1992) 85.}
\ref{schellekens89b}{B.~Schellekens, ed. \underbar{Superstring Construction},
 (North-Holland Physics, Amsterdam, 1989).}
\ref{schellekens89c}{B.~Schellekens, CERN-TH-5515/89.}

 \ref{schwarz87}{M.~Green, J.~Schwarz, and E.~Witten,
\underbar{Superstring Theory}, {\bf Vols. I \& II},
(Cambridge University Press, New York, 1987).}

\ref{turok87a}{D.~Mitchell and N.~Turok, {\it Nucl.~Phys.~}{\bf B294}
(1987) 1138.}
\ref{turok87b}{N.~Turok, Fermilab 87/215-A (1987).}

\ref{verlinde88}{E.~Verlinde, {\it Nucl.~Phys.~}{\bf B300}
(1988) 360.}

\ref{warner90}{N.~Warner, {\it Commun.~Math.~Phys.~}{\bf 130} (1990) 205.}

\ref{wilczek90} {F.~Wilczek, ed. \underbar {Fractional Statistics and Anyon
Superconductivity}, (World Scientific, Singaore, 1990) 11-16.}

\ref{wise84}{L.~Abbott and M.~Wise \NPB 224 (1984) 541.}

\ref{witten92}{E.~Witten, preprint IASSNS-HEP-93-3.}

\ref{vafa1}{R.~Brandenberger and C.~Vafa, {\it Nucl.~Phys.}
{\bf B316} (1989) 391.}
\ref{vafa2}{A.A.~Tseytlin and C.~Vafa, {\it Nucl.~Phys.}
{\bf B372} (1992) 443.}
\ref{vafa3}{C.~Vafa, private communication.}

\ref{zamol87}{A.~Zamolodchikov and V.~Fateev, {\it Sov.~Phys.~}JETP
{\bf 62} (1985)  215; {\it Teor.~}{\it Mat.}
{\it Phys.~}{\bf 71} (1987) 163.}
\end{putreferences}
\hfill\vfill\eject

{\parindent=.6cm
\n{\bf\bigfonts Figure Captions}\hfill\\
\vskip .2cm

\noindent Figure 1. Nine-dimensional torus (represented as
a nine-dimensional box with periodic bounday conditions) containing
winding mode strings around a single dimension
\vskip .2truecm

\noindent Figure 2. Number of steps required for collision to
occur between two winding mode strings,
with $98\%$ probability, for three compactified dimensions,
verses $R_0$, the radius of compactification at which the winding
modes have effectively fallen out of equilibrium.
($R_0$ in units of Planck length.)
\vskip .2truecm

\noindent Figure 3. Maximum expansion rate, $H$, allowed for collision to
occur between two winding mode strings,
with $98\%$ probability, for three compactified dimensions,
verses $R_0$, the radius of compactification at which the winding
modes have effectively fallen out of equilibrium.
($H$ in units of inverse Planck time and
$R_0$ in units of Planck length.)
\vskip .2truecm

\noindent Figure 4. Average dimension of decompactified
spacetime, $\langle D \rangle$,
for $H= 10^{-1}$ to $10^{-5}$,
verses $R_0$, the radius of compactification at which the winding
modes have effectively fallen out of equilibrium.
($H$ in units of inverse Planck time and
$R_0$ in units of Planck length.)
\vskip .2truecm

\noindent Figure 5. Histogram of the dimension of
decompactified spacetime, $D$,
for $H= 10^{-3}$ and varying $R_0$,
the radius of compactification at which the winding
modes have effectively fallen out of equilibrium.
($H$ in units of inverse Planck time and
$R_0$ in units of Planck length.)
50 trials run for each choice of $R_0$.
\vskip .2truecm

\noindent Figure 6. Average dimension of decompactified spacetime,
$\langle D \rangle$, for $H= 10^{-2}$ and $R_0 = 1.0$, verses
$c_0$, the constant of the effective string width
$\langle \omega \rangle= \sqrt{c_0 + {1\over 4\pi \sigma}\ln\,(l^2/\alpha')}
\,$.
($H$ in units of inverse Planck time and
$R_0,\, c_0$ in units of Planck length.)
\vskip .2truecm

\noindent Figure 7. Average dimension of decompactified spacetime,
$\langle D \rangle$, for $H= 10^{-3}$ and $R_0= 1.5$, verses
$R_{\rm max}$, the maximum radius obtainable for a given direction around
which two or more winding mode strings exist.
($H$ in units of inverse Planck time and
$R_0,\, R_{\rm max}$ in units of Planck length.)

\hfill\vfill\eject

\bye